\documentclass[twocolumn,amsmath,amssymb,prb]{revtex4-1}
\usepackage{graphicx}% Include figure files
\usepackage{dcolumn}% Align table columns on decimal point
\usepackage{multirow}
\usepackage{bm}% bold math
\usepackage{placeins}
\usepackage{color}

\begin{document}

\title{Structural properties of Silicon-Germanium and Germanium-Silicon Core-Shell Nanowires}

\author{Conn O'Rourke}
\affiliation{London Centre for Nanotechnology, University College London, 17-19 Gordon St, London, WC1H 0AH}
\affiliation{International Centre for Materials Nanoarchitectonics (MANA), National Institute for Materials Science (NIMS), 1-1 Namiki, Tsukuba, Ibaraki 305-0044, Japan}
\author{Shereif Y. Mujahed}
\affiliation{London Centre for Nanotechnology, University College London, 17-19 Gordon St, London, WC1H 0AH}
\affiliation{International Centre for Materials Nanoarchitectonics (MANA), National Institute for Materials Science (NIMS), 1-1 Namiki, Tsukuba, Ibaraki 305-0044, Japan}
\author{Chathurangi Kumarasinghe}%
\affiliation{London Centre for Nanotechnology, University College London, 17-19 Gordon St, London, WC1H 0AH}
\affiliation{International Centre for Materials Nanoarchitectonics (MANA), National Institute for Materials Science (NIMS), 1-1 Namiki, Tsukuba, Ibaraki 305-0044, Japan}
\author{Tsuyoshi Miyazaki}%
\affiliation{International Centre for Materials Nanoarchitectonics (MANA), National Institute for Materials Science (NIMS), 1-1 Namiki, Tsukuba, Ibaraki 305-0044, Japan}
\author{David R. Bowler}%
\email{david.bowler@ucl.ac.uk}
\affiliation{London Centre for Nanotechnology, University College London, 17-19 Gordon St, London, WC1H 0AH}
\affiliation{International Centre for Materials Nanoarchitectonics (MANA), National Institute for Materials Science (NIMS), 1-1 Namiki, Tsukuba, Ibaraki 305-0044, Japan}
\affiliation{Department of Physics \& Astronomy, University College London, Gower St, London, WC1E 6BT}%

\date{\today}

\begin{abstract}
Core-shell nanowires made of Si and Ge can be grown experimentally with excellent control for different sizes of both core and shell.  We have studied the structural properties of Si/Ge and Ge/Si core-shell nanowires aligned along the $[110]$ direction, with diameters up to 10.2~nm and varying core to shell ratios, using linear scaling Density Functional Theory (DFT). We show that Vegard's law, which is often used to predict the axial lattice constant, can lead to an error of up to 1\%, underlining the need for a detailed \emph{ab initio} atomistic treatment of the nanowire structure. We analyse the character of the intrinsic strain distribution and show that, regardless of the composition or bond direction, the Si core or shell always expands. In contrast, the strain patterns in the Ge shell or core are highly sensitive to the location,  composition and bond direction.  The highest strains are found at heterojunction interfaces and the surfaces of the nanowires.  This detailed understanding of the atomistic structure and strain paves the way for studies of the electronic properties of core-shell nanowires and investigations of doping and structure defects.
\end{abstract}


\maketitle

\section{\label{Intro}Introduction}

Scaling down the size of the current generation of electronic devices has led to an increased interest in 
semiconductor nanostructures, such as nanowires and 
nanotubes \cite{Lu:2007ef,Thelander:2006sw,Avouris:2002te,Yang:2005yc,Goldberger:2006cn}. 
Quantum size effects and high surface to volume ratios in these structures as a result of one or more reduced dimensions can lead to highly tunable and unique electronic, optical and transport properties, which have the potential to be exploited in next generation electronic devices\cite{Lu:2005fu,Xiang:2006mi,Zhao:2004vk}. Recently, Si/Ge core-shell nanowires have been studied extensively, both experimentally and theoretically since they are promising candidates for such applications with the valence band offset between Ge and Si offering a unique opportunity to control spacial carrier confinement and carrier transport\cite{Lu:2005fu,Xiang:2006mi,Fukata:2012xm}.  In this work, we use linear scaling DFT to study the structural and strain properties of Si/Ge and Ge/Si core-shell nanowires, as a function of nanowire composition and diameter, from $\sim$5\,nm to $\sim$10\,nm.

Pure Si and Ge nanowires are typically grown by chemical vapour deposition within the VLS method\cite{Dasgupta:2014nk}, and this approach affords a significant amount of control during the deposition process; the smallest nanowires grown by this method vary between 1nm and 7nm in diameter\cite{Ma:2003qj,Wu:2004un}. As well as pure nanowires, it is possible to create core-shell heterostructures\cite{Lauhon:2002eg} using simple CVD after the VLS growth of a nanowire; these can be created with a Si core and Ge shell (and vice-versa), simply by changing the growth conditions and reactants during the deposition process\cite{Xiang:2006mi,Lauhon:2002eg}.  Shells as thin as 2--3nm have been grown\cite{Nguyen:2014ms,Noguchi:2015ir}, though on cores which are a little larger than those studied here (15--20nm).

\begin{figure}[h]
  \begin{center}
    \includegraphics[clip, width=0.48\textwidth]{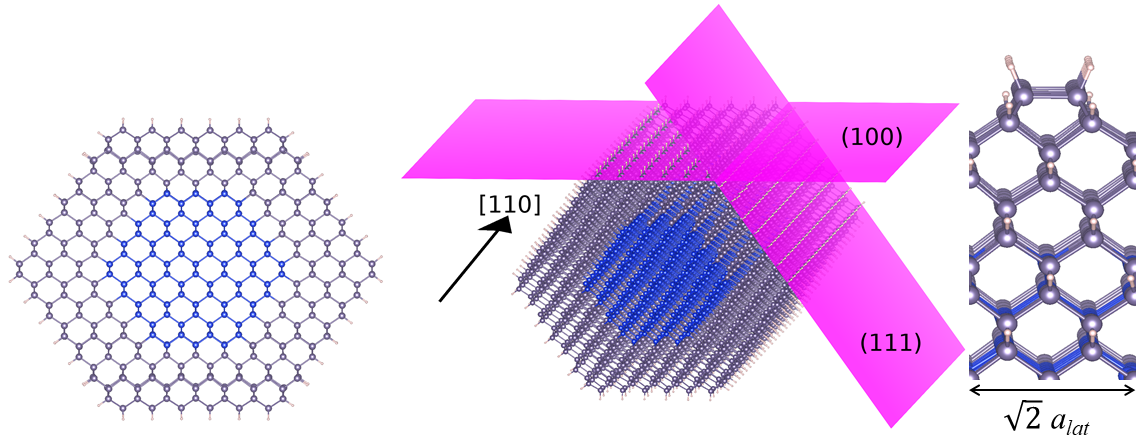}
    \caption{Si core, Ge shell nanowire model illustrating exposed surfaces and growth direction, along with a side view of the passivated $2\times1$ surface reconstruction of the $(100)$ surface of the SiGe-NW. \label{NW_model}}
   \end{center}
\end{figure}

VLS grown Si nanowires are known to form essentially exclusively along the $\langle 110\rangle$ direction\cite{Ma:2003qj,Wu:2004un}
for diameters up to 10\,nm, exposing the $(100)$ and $(111)$ surfaces\cite{Ma:2003qj} (beyond this diameter, they transition to a $\langle 111\rangle$ direction, with shapes that tend towards round; ).  It has also been shown that growth of shells on these small diameter nanowires leads to smooth, dislocation-free shells\cite{Goldthorpe:2009rs}.  We have therefore modelled the core-shell
nanowire system for diameters up to this cross-over point, using the model depicted in Fig.~\ref{NW_model}, for both Si-core and Ge-core nanowires. In all cases, the surfaces have been passivated using hydrogen atoms.

% STRAIN
Strain has a significant impact on the structural stability as well as electronic properties of nanostructures\cite{Bhushan:2014ta,Jaishi:2017dk,Yang:2008kl}. The bonds in Si/Ge core-shell nanowires are intrinsically strained, primarily due to the lattice mismatch of around 4\%  between Si and Ge, though there are also strains due to surface reconstructions. The nature of these anisotropic  strain patterns will be determined by several factors including shape and composition of the NWs, characteristics of the heterostructure interface, the amount of lattice mismatch and the elastic parameters of the constituent materials\cite{Peng:2011aw,Larsson:2006ti,Liu:2011ub}. 
Detailed analysis of the local strain distribution will enable us to identify highly compressed and extended areas in the nanowires, where dislocations and other defects are most likely form to relieve the strain. This will also inform studies of where dopants are stable in such nanowires,  since it is known that certain dopants such as As tend to be more stable in areas under tensile strain, while others such as P are known to be stable under compressive strain\cite{Dunham:2006sk}.

Strained areas, particulary near the surfaces, can have a significant effect on chemical reactivity of nanowires because strain can lift degeneracy and cause band splitting\cite{Ciraci:1988eo,Van-de-Walle:1986ca}, changing the electronic structure. This can alter reaction rates or allow reactions that would otherwise not proceed. Precise knowledge of the strain patterns will help to identify local changes in electronic structure and hence determine how strain will affect reaction rates near the surfaces. Further, it will help to predict spatial variation of electron transport and optical properties in nanowires. Even though the intrinsic strain has been demonstrated to critically affect both structural and electronic properties, it has not been given  much attention in the literature, especially when it comes to relatively large nanowires with diameters over 5~nm (previous studies have maximum sizes of 4.7\,nm\cite{Peng:2011aw}). 

% AXIS
The axial lattice parameter, which is sensitive to the changing of the NW composition, is a key input to any core-shell nanowire atomistic or electronic structure computation, and has a significant effect on the structure and hence electronic properties of the nanowire. Determining the optimum axial lattice parameter using first principle methods can be time consuming, so an estimate based on Vegard's law \cite{Vegard:1921vu,Denton:1991gv} is often used as a first approximation, though is not always accurate \cite{Nduwimana:2008xt}. Vegard's law determines the lattice parameter of a solid solution based on lattice parameters of the pure components and their relative concentrations in the solution\cite{Jacob:2007sd}. The law is purely empirical, and was first derived in the context of solid ionic solutions. Deviations from the linear behaviour assumed in Vegard's law are often observed for other materials\cite{Jacob:2007sd,West:2007fs}.
In situations where the lattice parameter of a solid solution is known experimentally (e.g. through diffraction data), Vegard's law is often used effectively in reverse, to estimate the relative compositions\cite{Nishiyama:1990uo}. It has been shown that the lattice constant of $\text{Si}_{1-x}\text{Ge}_x$ \emph{alloy} nanowires can be well predicted by means of Vegard’s law\cite{Iori:2014fm}, in line with bulk and thin film alloys.  Here, we use linear scaling DFT to explore the applicability of Vegard's law estimates to Si/Ge core-shell nanowires, even though these heterostructures do not fall within the area where the law was first derived. This analysis is needed for core-shell nanowires, particularly due to highly inhomogeneous and anistropic strain distribution generated as a result of non-statistical distribution of Si and Ge atoms; however, it is lacking in the literature, particularly for nanowires with diameters over 5~nm\cite{Musin:2005zv,Nduwimana:2008xt}. 
 
Ab initio approaches based on DFT are commonly used in studiying these structural factors of nanostuctures. Due to computational demands, the overwhelming majority of the work based on the DFT framework has examined NWs with diameters of around 5~nm or less\cite{Amato:2011xf,Yang:2008kl,Musin:2006ao,Amato:2009uc,Amato:2009pi,Musin:2005zv,Migas:2007qr,Peng:2010gk}. Given that the dimensions of experimentally studied core-shell nanowires are typically of the order of tens of nanometers, it is clear that significantly larger systems must be examined theoretically. Recent implementations of DFT methods which scale linearly with system size\cite{Bowler:2012nu,Bowler:2010pm,Bowler:2002bj,Skylaris:2008xz,Soler:2002kn} have made accurate modeling of such larger scale systems possible with reasonable computational cost, thereby affording a much better insight into the properties of core-shell nanowires of physically realistic size. 

In this work, we study the relationship between structural properties and overall composition of  Si/Ge and Ge/Si core-shell nanowires with diameters in the range 4.9---10.2~ nm using the linear scaling DFT code \textsc{Conquest}. We explore the accuracy of Vegard's law in determining the axial lattice constant, and investigate the intrinsic strain patterns of the core-shell nanowires, particularly how these change with core to shell ratio, core-shell composition and diameter. 

\section{\label{CompDeet}Approach}
All the calculations of nanowire structure used the linear scaling DFT code, \textsc{Conquest}\cite{Bowler:2002bj}, using the PBE GGA functional\cite{Perdew1996}.  \textsc{Conquest} is a linear-scaling, or $\mathcal{O}(N)$ DFT code with the capability to perform first-principles DFT calculations on systems of up to millions of atoms\cite{Nakata:2017be,Bowler:2012nu}. Since the details of implementation of \textsc{Conquest} have been discussed elsewhere\cite{Bowler:2010pm,Bowler:2012nu,Bowler:2002bj}, we summarize only the main principles needed to explain the current approach.

It is well known that the DFT ground state can be obtained by minimising the total energy with respect to the Kohn-Sham (KS) density matrix $\rho(r,r')$, which is formally defined as,
\begin{equation}
\rho(r,r')=\sum_n f_n \Psi_n(r) \Psi_n(r'),
\end{equation}
where $\Psi_n$ and $f_n$ are the $n^{th}$ KS orbital and its occupation number, respectively. 

In \textsc{Conquest}, $\rho(r,r')$ is represented in terms of localized orbitals $\phi$ centered on the atoms, known as `support functions':
\begin{equation}
\rho(r,r')=\sum_{i\alpha,j\beta} \phi_{i\alpha}(r) K_{i\alpha,j\beta}\phi_{j\beta}(r').
\end{equation}
where $i$($j$) indicates an atom and $\alpha$($\beta$) runs over the support functions on the $i^{th}$($j^{th}$) atom. While the support functions themselves can be represented in terms of basis functions, we use a one-to-one mapping between support functions and pseudo-atomic orbitals (PAOs)\cite{Torralba:2008wm}. For all our calculations we have employed a single-$\zeta$ plus polarisation orbital basis of PAOs. This basis has been chosen to give a balance between accuracy and speed, and is the largest basis that can be easily used with linear scaling.  The PAO cutoffs are chosen to give an optimised bulk lattice paramter within 1\% of the experimental values for both Si and Ge respectively. 

The coefficients $K_{i\alpha,j\beta}$ are the density matrix elements in the basis of support functions. In \textsc{Conquest}, the density matrix ($K$) can either be calculated by the conventional direct diagonalization with $\mathcal{O}(N^3)$ scaling or by using the density matrix minimization method proposed by Li, Nunes and Vanderbilt (LNV)\cite{Li:1993lg} with $\mathcal{O}(N)$ scaling. In the LNV method, which is used for our calculations, K is expressed in terms of an auxiliary density matrix $L$ by the matrix relation:
\begin{equation}
K = 3LSL-2LSLSL 
\end{equation}
where $S_{i\alpha,j\beta} = \langle\phi_{i\alpha} \mid \phi_{j\beta}\rangle$ is the overlap matrix of support functions. To achieve linear scaling with the LNV method, a spatial cut-off $R_L$ must be imposed on the $L$-matrix so that its value is zero when the distance between the centres of the support-functions exceeds $R_L$. Imposing this spatial cutoff is justified by the the nearsightedness of electronic matter\cite{Kohn:1995mq} and needed to ensure O(N) scaling. The $R_L$ value is a compromise between the accuracy and the computational cost, with larger $R_L$ leading to better accuracy at a higher computational cost. For all our calculations, a spatial cut-off $R_L = 16 a_0$ was selected, at which $\mathcal{O}(N)$ forces in the system were converged to within the force tolerance of the exact diagonalisation results for Si and Ge surfaces.\cite{Miyazaki:2007kr}

Numerical integration of PAOs in space is required to form the local part of the Hamiltonian matrix in \textsc{Conquest}. This integration takes place on an integration grid and its spacing will contribute to determining the overall accuracy of the calculations. The accuracy can be improved by using a finer grid, i.e. increasing the integration grid cutoff, however this leads to a rapid increase in the computational costs. For the inputs including pseudopotentials and basis functions used in our calculations, an integration grid cut-off of 100 Ha has been identified as sufficient for the DFT energy to converge.  

The nanowires present an interesting challenge for structural optimisation: the nanowire is constrained along the $[110]$ axial direction, but is free along the radial direction.  We perform a two-stage relaxation to find the optimal axial lattice parameter: a value for the lattice parameter is chosen, and we perform a structural relaxation of the nanowire.  This process is repeated for different values of the axial lattice parameter, $a_\text{lat}$, to find the lowest energy value.
The simulation cell includes $\sim$13\AA\ vacuum in lateral x- and y-directions to avoid any interactions between the images of neighboring nanowires from periodicity. 
Structural optimisations have been performed using the FIRE algorithm\cite{Bitzek:2006wo}, 
until force components on each atom were less than 0.0005 $\text{Ha}/a_0$.

\begin{figure}[h]
  \begin{center}
   \begin{tabular}{c c c}
     \begin{minipage}[h]{0.1\textwidth}
        \includegraphics[trim = 0mm 0mm 0mm 0mm, clip, width=1.\textwidth]{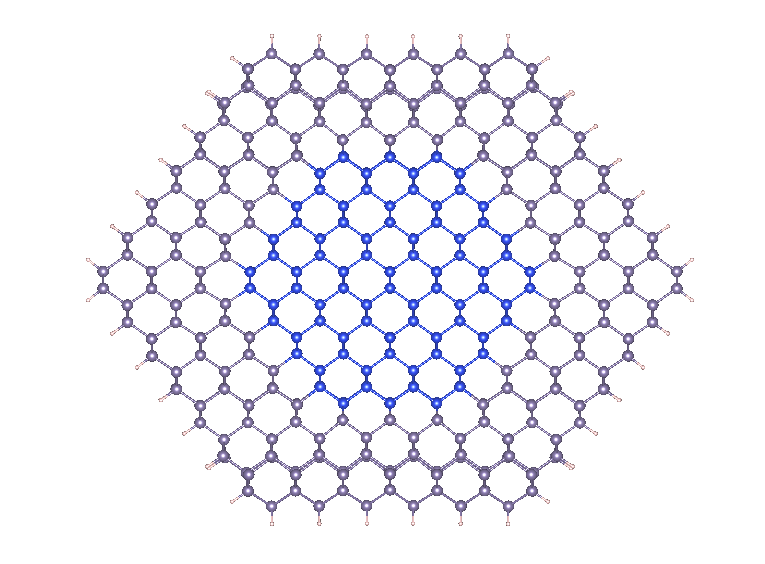}
      \end{minipage}&
      \begin{minipage}[h]{0.125\textwidth}
        \includegraphics[trim = 0mm 0mm 0mm 0mm, clip, width=1.\textwidth]{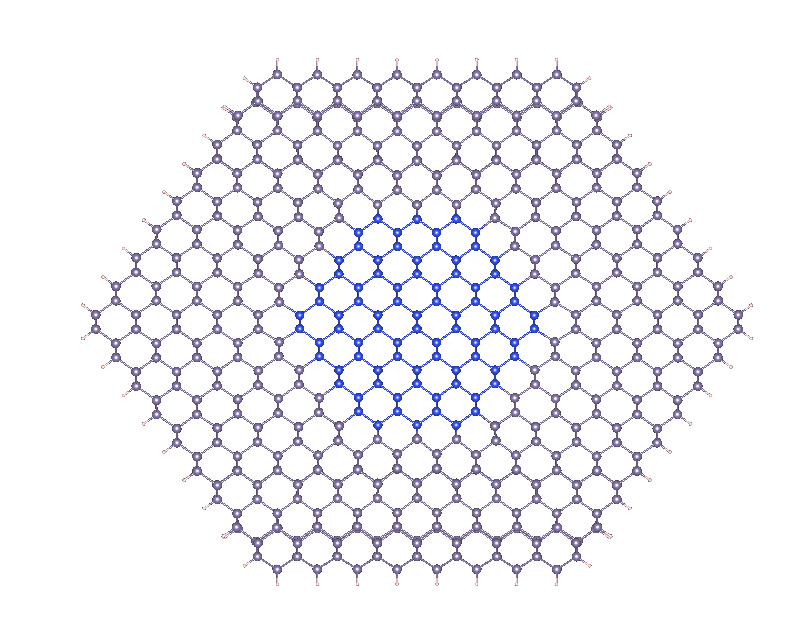}
      \end{minipage}&
      \begin{minipage}[h]{0.15\textwidth}
        \includegraphics[trim = 0mm 0mm 0mm 0mm, clip, width=1.\textwidth]{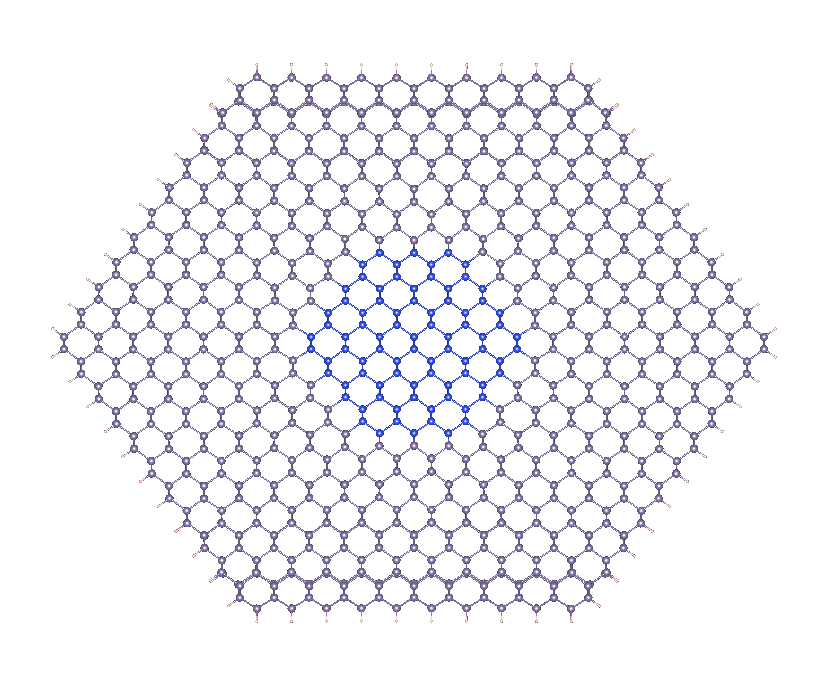}
      \end{minipage}\\
      3\_3 & 3\_5 & 3\_7\\
           \begin{minipage}[h]{0.13\textwidth}
        \includegraphics[trim = 0mm 0mm 0mm 0mm, clip, width=1.\textwidth]{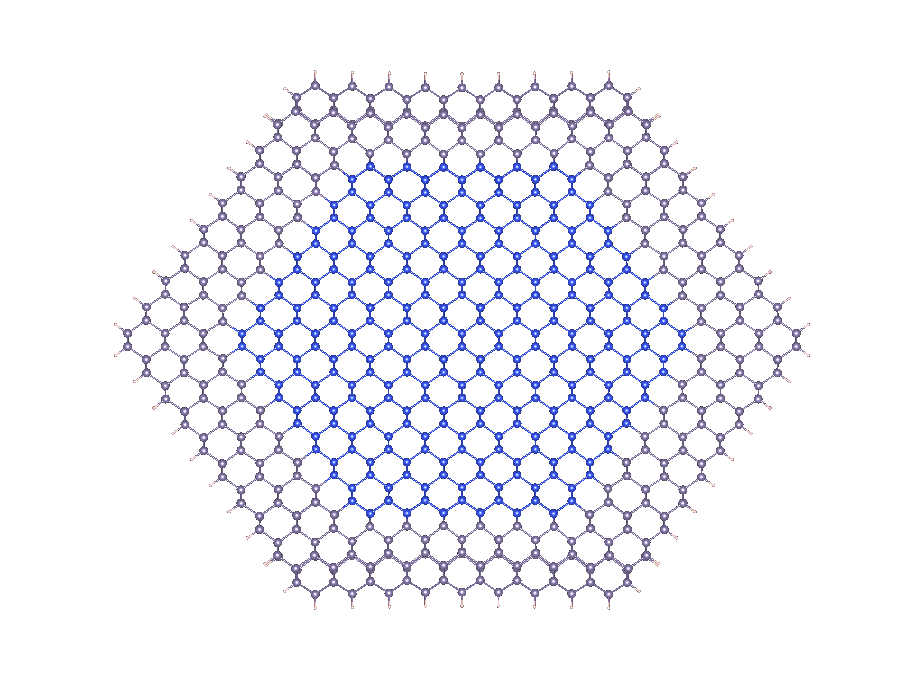}
      \end{minipage}&
      \begin{minipage}[h]{0.155\textwidth}
        \includegraphics[trim = 0mm 0mm 0mm 0mm, clip, width=1.\textwidth]{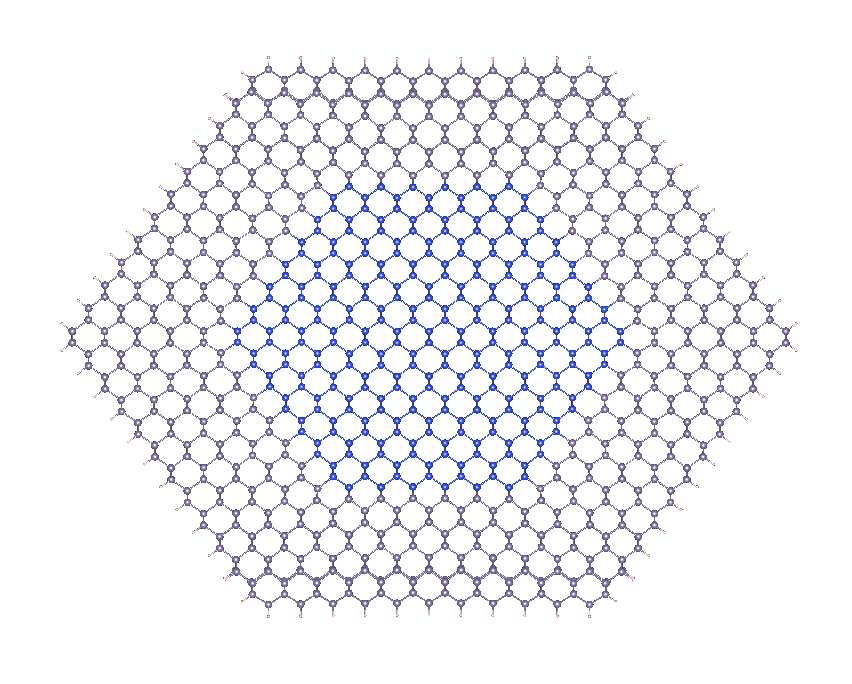}
      \end{minipage}&
      \begin{minipage}[h]{0.195\textwidth}
        \includegraphics[trim = 0mm 0mm 0mm 0mm, clip, width=1.\textwidth]{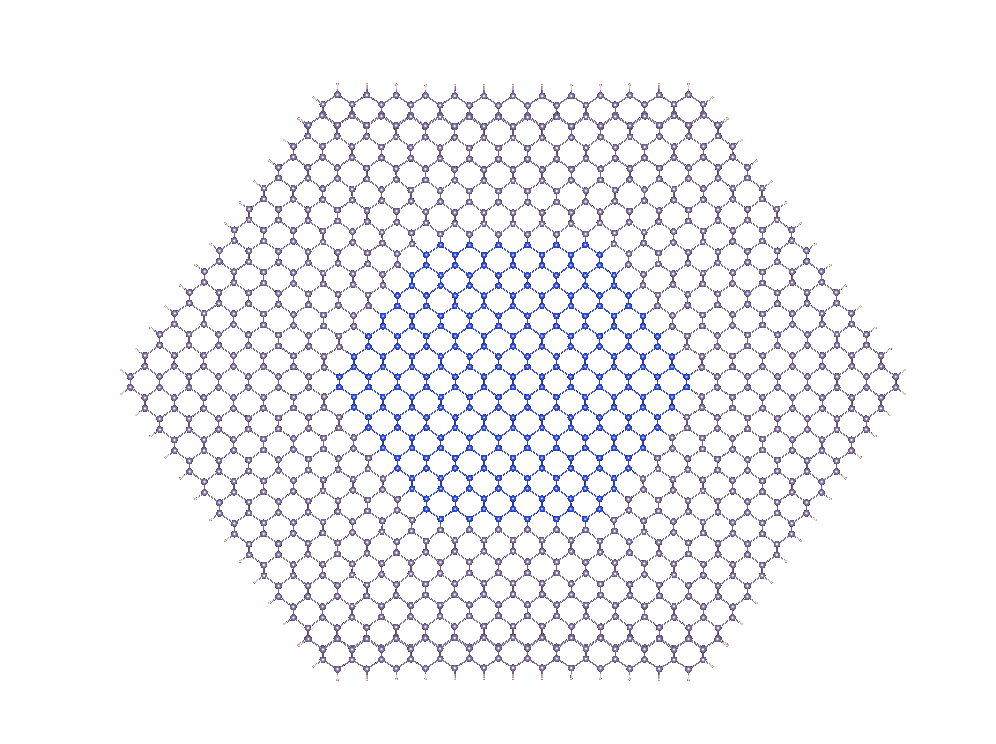}
      \end{minipage}\\
      6\_3 & 6\_5 & 6\_7\\ 
        \end{tabular}
     \caption{SiGe-NW models used throughout, labelled C\_S where the index C represents the number of layers in the core and S the surface. Shell thickness increases left to right, and core thickness top to bottom. (The same motifs have been used for the GeSi-NW models).}
    \label{core_shell}
   \end{center}
\end{figure}

Our intention is to examine how the relative core-shell thicknesses will affect the structural properties
of these nanowires, and as such we have performed calculations varying independently the shell and core size   
for both Si-core Ge-shell (SiGe-NWs) and Ge-core Si-shell nanowires (GeSi-NWs).  The shape of the core is set by the free energies of the surfaces\cite{Ma:2003qj,Wu:2004un}, and we have chosen to maintain the shape in the shell for simplicity.
Cross-sections of the models examined for the SiGe-NW variants can be seen in Fig.~\ref{core_shell}, with
the same motifs used for the GeSi-NWs. The smallest of our models is approximately 4.9~nm in diameter, 
containing 612 atoms, and the largest is approximately 10.2~nm in diameter and contains 2404 atoms.

\begin{center}
\renewcommand{\arraystretch}{1.2}
\begin{table*}[t]
%\begin{center}
\begin{minipage}{0.95\textwidth}
\begin{center}
\begin{tabular}{c c c c c c c c}
\hline
\hline
& &\multicolumn{6}{c}{Model} \\
                          &                & 3\_3 & 3\_5 & 3\_7 & 6\_3 & 6\_5 & 6\_7 \\
\hline
\hline
\multirow{ 2}{*}{SiGe-NW} & $a^{min}_{lat}$& 5.593& 5.619& 5.624& 5.574& 5.532& 5.565\\
                          & $a^{lin}_{lat}$& 5.590& 5.628& 5.631& 5.549& 5.584& 5.604\\
                            & $error(\%)$& 0.054& -0.160& -0.125& 0.449& -0.940& -0.701\\
\hline
\multirow{ 2}{*}{GeSi-NW} & $a^{min}_{lat}$& 5.531& 5.475& 5.460& 5.574& 5.554& 5.492\\
                          & $a^{lin}_{lat}$& 5.500& 5.471& 5.457& 5.539& 5.505& 5.485\\
                          & $error(\%)$& 0.561&	0.073& 	0.055& 	0.628& 	0.882	& 0.128\\
\hline
\hline
\end{tabular}
\end{center}
\end{minipage}
%\normalsize
\caption{Calculated axial lattice parameters using DFT, $a^{min}_{lat}$, for SiGe model nanowires depicted in Fig.~\protect\ref{core_shell}, along with the interpolated axial lattice parameter, $a^{lin}_{lat}$ based upon Eq.~\protect\ref{linear_interp}. Error is the percentatge of deviation of $a^{lin}_{lat}$ from $a^{min}_{lat}$.
All measurements are given in \AA.}
%\end{center}
\label{SiGelat}
\end{table*}
\renewcommand{\arraystretch}{1.0}
\end{center}

\section{Results: Calculated Axial Lattice Parameters}

One approach for calculating the \emph{ideal} axial lattice parameter is to treat the nanowire as a solid solution, and to use Vegard's law\cite{Vegard:1921vu,Denton:1991gv}, taking a linear interpolation based on the relative numbers of Si and Ge atoms as follows:

\begin{equation}
a^{lin}_{\text{lat}}=\frac{N^{Si}a_{\text{lat}}^{Si}+N^{Ge} a_{\text{lat}}^{Ge}}{N^{Si}+N^{Ge}},
\label{linear_interp}
\end{equation}

where $N^{Si}$ and $N^{Ge}$ are the total number of Si and Ge atoms present in the nanowire respectively. The bulk Si and Ge  lattice constamts, $a_{\text{lat}}^{Si}$  and $a_{\text{lat}}^{Ge}$ are taken as $5.432$\AA\ and $5.658$\AA\ \cite{Dargys:1994cz} respectively, for all calculations.

Another approach is to plot DFT calculated total system energy vs overall lattice parameter($a_{\text{lat}}$) by varying the lattice parameter explicitly, and fitting to find the optimum lattice parameter corresponding to the minimum energy structure. This lattice parameter corresponding to the minimum energy structure is the \emph{ideal} axial lattice parameter($a_{\text{lat}}$) used for fixing the axial atomic distances and axial simulation cell length in structural relaxations performed in the next sections.

\begin{figure}[h]
  \begin{center}
    \begin{minipage}[h]{0.45\textwidth}
        \rotatebox[origin=c]{0}{\includegraphics[angle=-90,origin=c,trim = 30mm 20mm 20mm 30mm, clip, width=1.\textwidth]{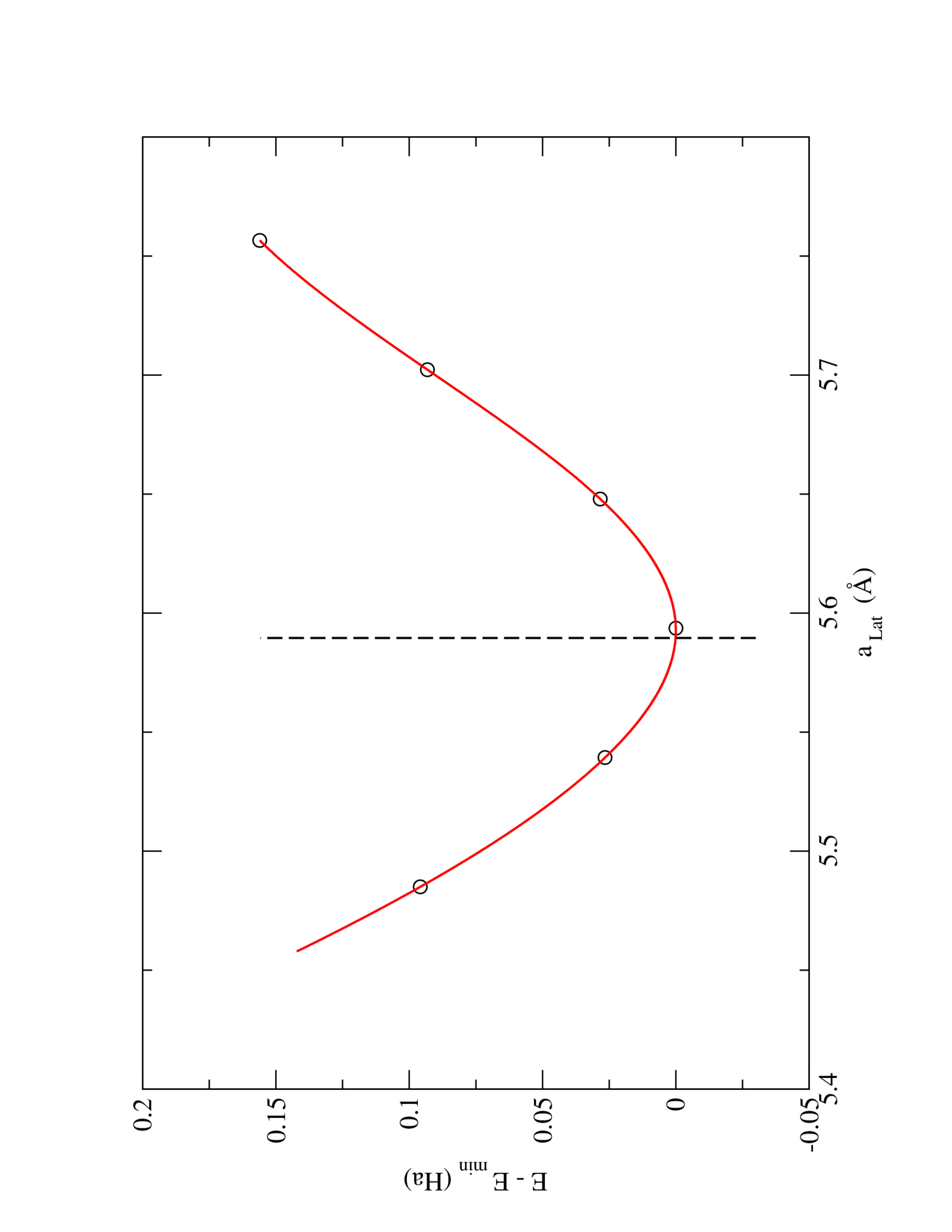}}
      \end{minipage}
   \caption{Energy variation with axial lattice parameter for the 3\_3 SiGe-NW. The red line is a cubic spline fitted
to the data, and the black dotted line represents the axial lattice parameter calculated analytically by eq. \eqref{linear_interp} }
  \label{3_3_lat}
  \end{center}
\end{figure}

\begin{figure}[b]
   \begin{center}
     \begin{minipage}[h]{0.3\textwidth}
        \includegraphics[trim = 0mm 0mm 0mm 0mm, clip, width=1.0\textwidth]{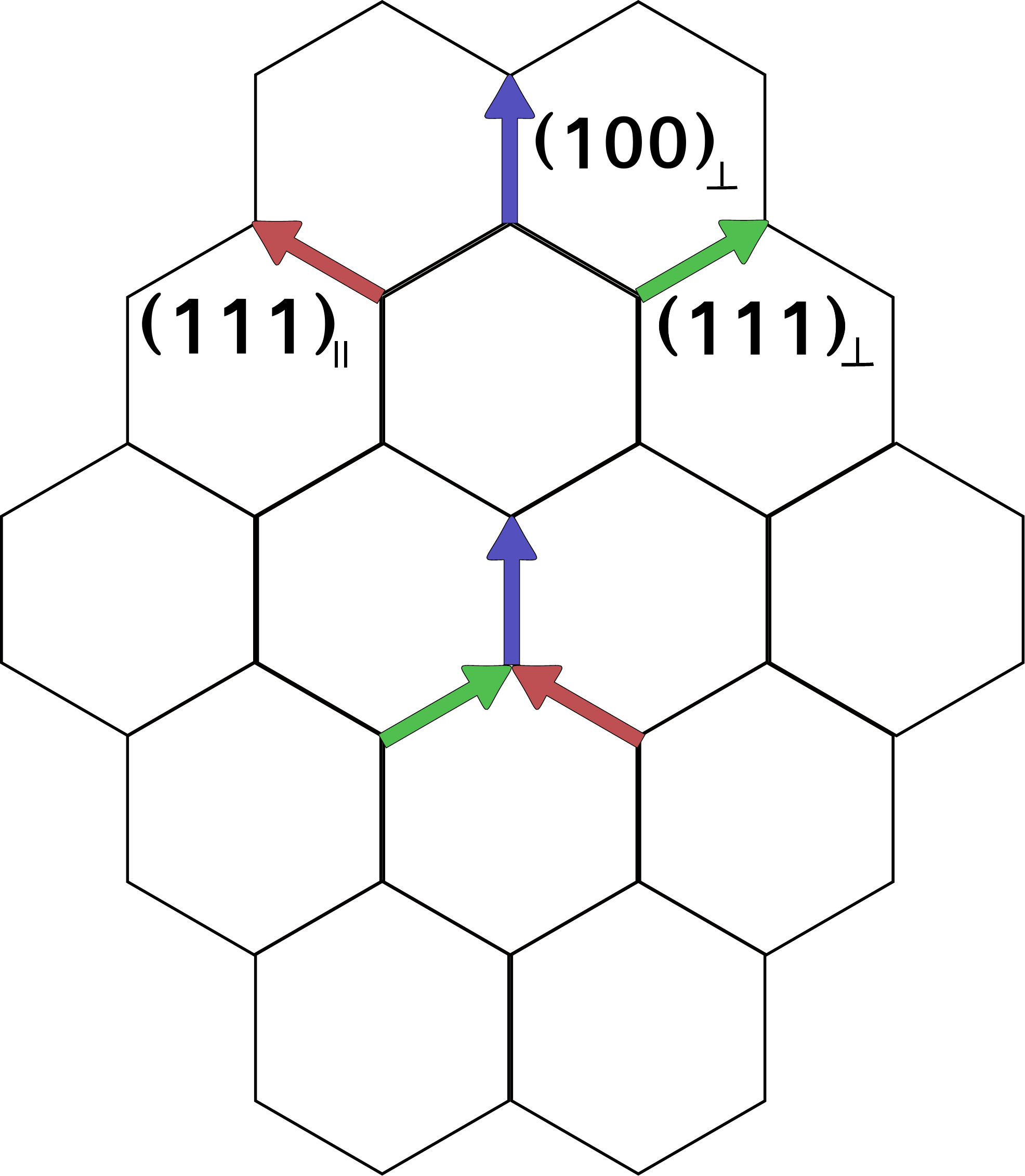}
      \end{minipage}
    \caption{Bond labelling: Cross sectional schematic of the NW models used, with three different bond types labelled, $(100)_\perp$ in blue, $(111)_\perp$ in green and $(111)_\parallel$ in red. }
    \label{BondLabel}
  \end{center}
\end{figure}

%New figure
\begin{figure*}[t]
  \begin{center}
    \begin{tabular}{c c} % LHS: bonds RHS: grid
      \begin{minipage}[h]{0.4\textwidth}
        \includegraphics[trim = 0mm 0mm 0mm 0mm, clip, width=\textwidth]{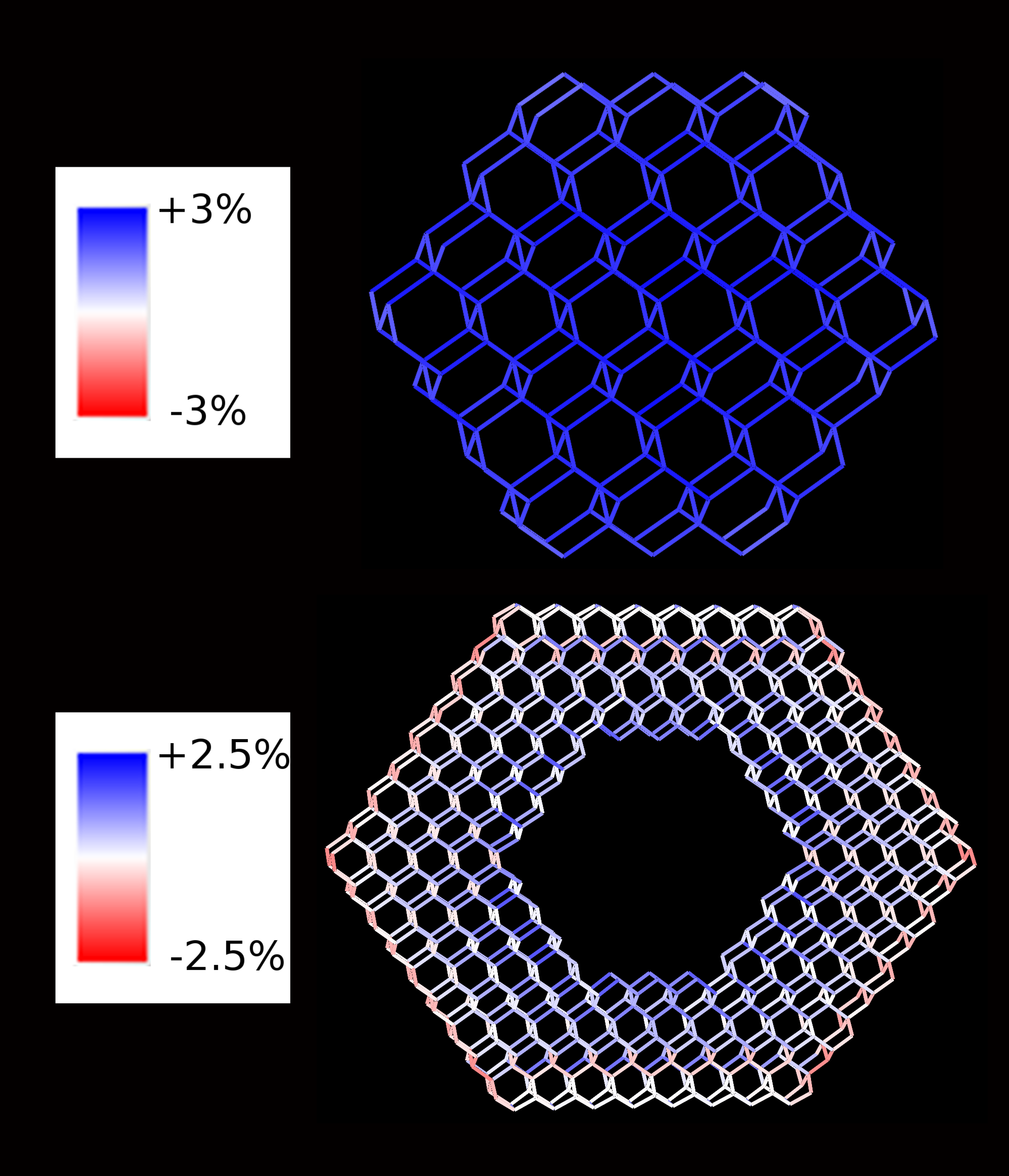}
      \end{minipage}&
      \begin{tabular}{c c} % LHS: strain maps RHS: label
        \begin{tabular}{c c} % Columns of strain maps
          \begin{minipage}[h]{0.2\textwidth}
            \includegraphics[trim = 45mm 45mm 55mm 45mm, clip, width=1.\textwidth]{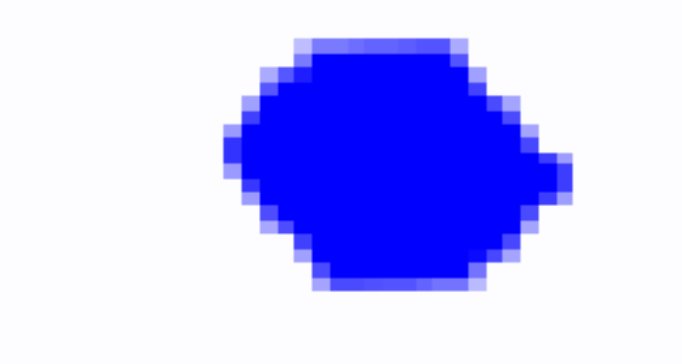}
          \end{minipage}&
          \begin{minipage}[h]{0.2\textwidth}
            \includegraphics[trim = 45mm 45mm 55mm 45mm, clip, width=1.\textwidth]{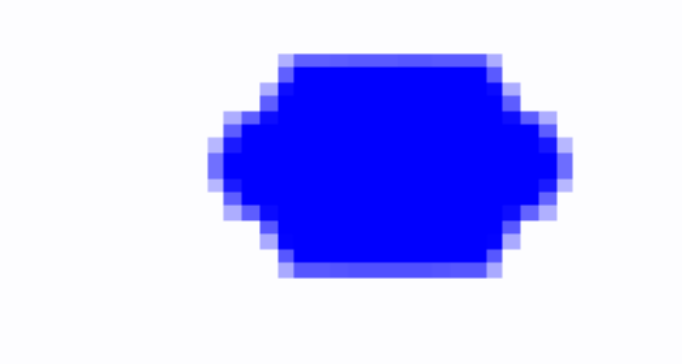}
          \end{minipage}\\
          \begin{minipage}[h]{0.2\textwidth}
            \includegraphics[trim = 30mm 30mm 40mm 30mm, clip, width=1.\textwidth]{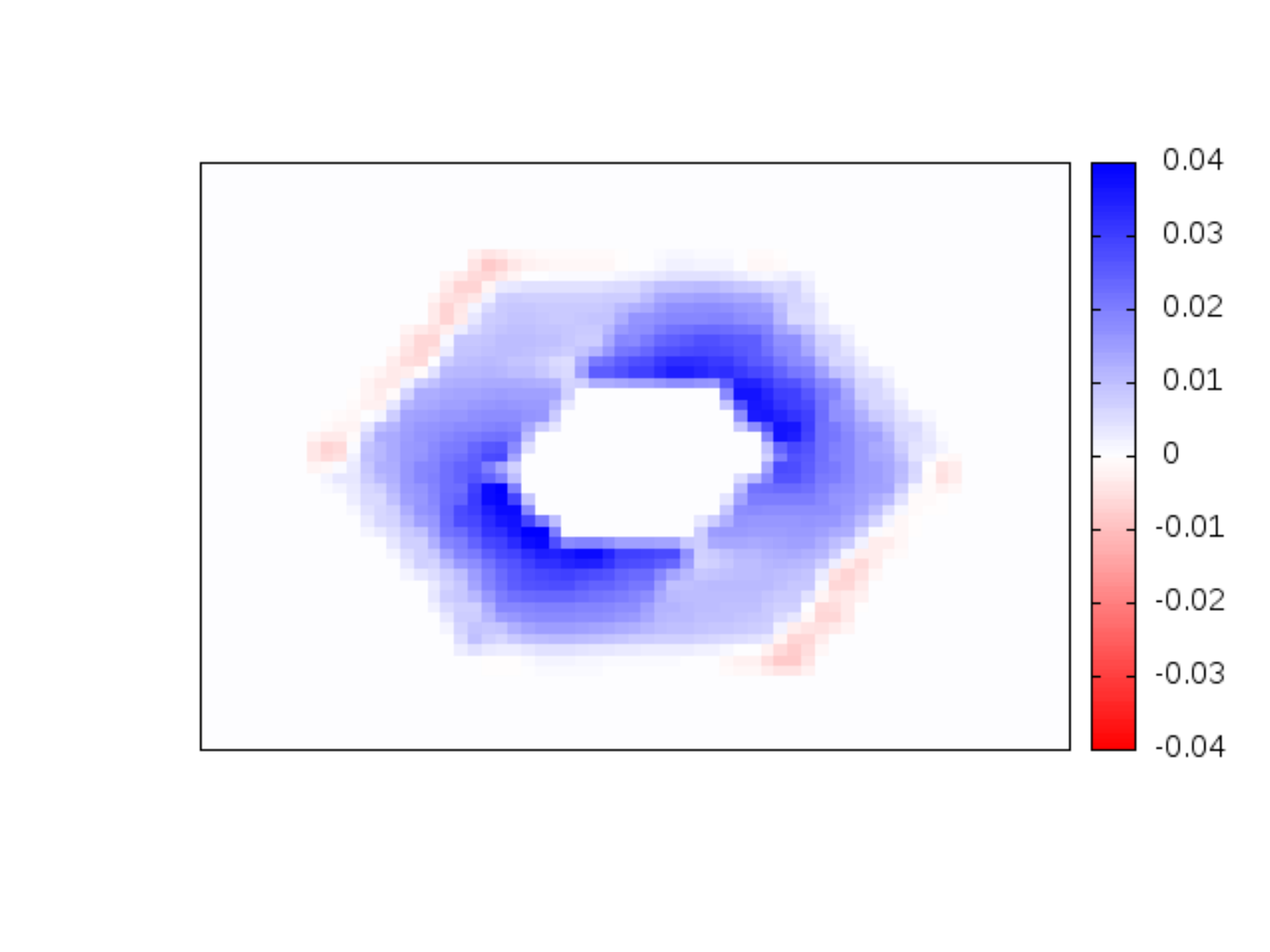}
          \end{minipage}&
          \begin{minipage}[h]{0.2\textwidth}
            \includegraphics[trim = 30mm 30mm 40mm 30mm, clip, width=1.\textwidth]{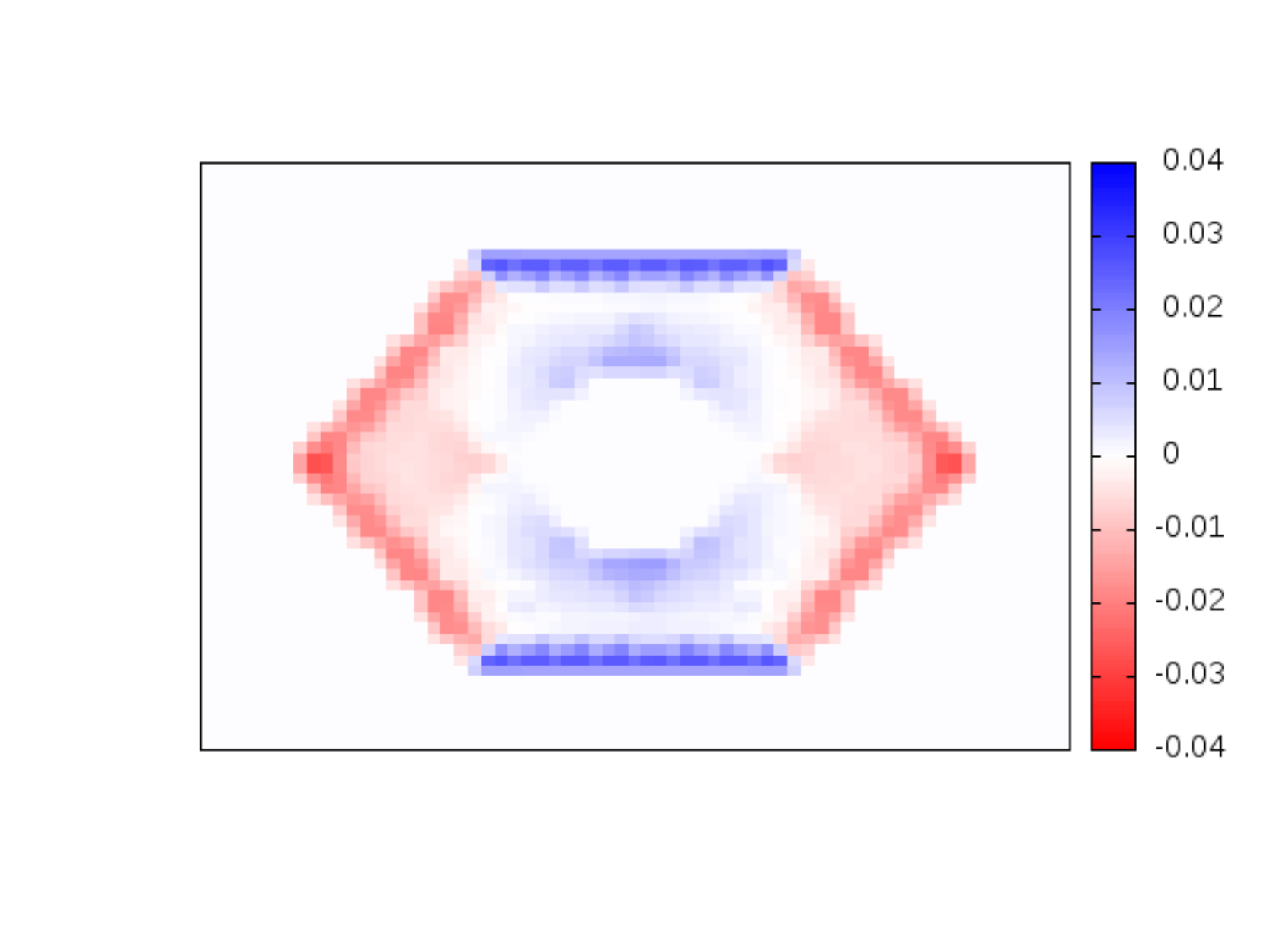}
          \end{minipage}\\
        $(\mathbf{111})_\parallel$&$(\mathbf{100})_\perp$
        \end{tabular}&
        \begin{minipage}[h]{0.05\textwidth}
          \includegraphics[trim = 0mm 0mm 0mm 0mm, clip, width=1.\textwidth]{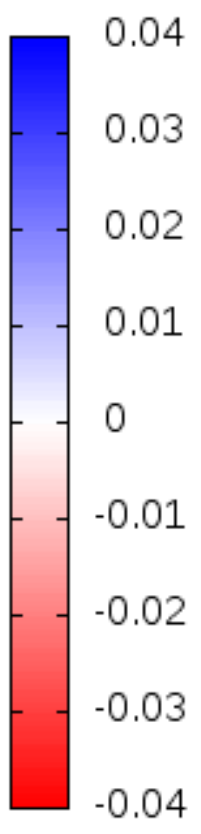}
        \end{minipage}
      \end{tabular}
    \end{tabular}
     \caption{Left: Percentage variation in bond length, relative to bulk bond lengths for the Si-Si bonds in the core (top), and Ge-Ge bonds in the shell (bottom) for SiGe-NWs with a 3-layer core and 5-layer Ge shell thickness. Right: Average bond strain map for the cross-section of the Si core (top) and Ge shell (bottom) of the same 3\_5 SiGe-NW. Maps are shown for bonds types $(111)_\parallel$ and $(100)_\perp$ (see Fig.~\protect\ref{BondLabel}); strains in $(111)_{\perp}$ are identical to those in $(111)_\parallel$ except reflected about the $(100)$ axis, exactly as the arrows shown in Fig.~\protect\ref{BondLabel}). }
    \label{Si3Ge_core}
   \end{center}
\end{figure*}

The effect of varying the 3\_3 SiGe-NW lattice parameter can be
seen in Fig.~\ref{3_3_lat}, along with the lattice parameter as calculated from Eq.~\ref{linear_interp}.
From our result for the 3\_3 SiGe nanowire we can see that, in this case, a linear interpolation gives a good approximation, with the minimum of the curve fitted to the data residing extremely
close to the interpolated point.

Results from carrying this procedure out for the rest of the SiGe and GeSi nanowires are given in Table~\ref{SiGelat}, along with the lattice parameter calculated via Eq.~\ref{linear_interp}. We can see that, in general, a linear interpolation using Eq.~\ref{linear_interp} provides a reasonable prediction of this \emph{ideal} axial lattice parameter, although far from perfect.  In a heteroepitaxial system such as this, with an intrinsic strain of $\sim$4\%, a further strain from an approximate lattice constant will have a considerable effect. 

For the GeSi-NWs we can see that the \emph{interpolated} axial lattice parameter consistently decreases towards the Si bulk lattice parameter with increasing shell size, as the number of Si atoms in the shell increases relative to the number of Ge atoms in the core.  For the optimised calculations, the same trend is seen, though the values are all larger than the interpolated values.

The interpolated axial lattice parameter of the SiGe-NWs increases with shell size as the system tends towards the Ge bulk lattice parameter, though this behaviour is not seen in the optimised parameters.  With a small silicon core, the lattice parameter follows the interpolated value reasonably; with the larger silicon core, the behaviour is quite different, and the lattice constant behaves non-monotonically, with large differences to the interpolated values.  Previous calculations of cylindrical core-shell NWs have also shown diameter dependent behaviour: very small ($\sim$1.5nm diameter) NWs show non-linear behaviour\cite{Musin:2005zv}, though the behaviour was monotonic; slightly large NWs (up to 4nm diameter)\cite{Peng:2011aw,Peng:2010gk} show behaviour similar to our large hexagonal NWs---non-monotonic for cylindical SiGe-NWs, with GeSi-NWs behaving monotonically. 

This non-monotonic departure from the simple linear interpolation of Vegard's law highlights an important point: while it is desirable to understand the structural properties of these nanowires in the simplest possible terms, the many different interfaces, coupled with the elastic anisotropy of both Si and Ge, provide an extremely complex system which will require careful, first principles simulation to explore fully.  The differences in lattice parameter found here (up to 1\%), and the departure from simple, expected behaviour, may give significant deviations in atomic and electronic structure which can only be fully explored using a technique such as linear scaling DFT, which can reach realistic simulation cell sizes.  This ability will help to understand systems in the field of next generation electronics, allowing systems of physically realistic size to be examined and trends exposed.

\section{Results: Intrinsic Strain}

Having calculated the axial lattice parameter corresponding to the minimum energy for each of our nanowire models, we now proceed to analyse the relaxed structures for this axial lattice in each case. We can see from Fig.~\ref{NW_model} that we have four $(111)$ surfaces in our hexagonal nanowire, and two $(100)$ surfaces. The (111) surface is very simple, without reconstruction, and is unlikely to have any effect beyond the influence of the surface as a boundary, while the (100) is more complex with Si-Si dimerisation along the [110] direction, typically leading to strained bonds and changes in bond angle.

We consider the directions of bonds in the nanowires carefully, as there is considerable anisotropy in the system, coming both from elastic anisotropy and from the different boundary conditions.  Looking at the cross sectional schematic in figure~\ref{BondLabel}, we see that in essence there are three orientations for the bonds in our nanowire model. First, bonds that have a significant vector component perpendicular to the $(100)$ surface
along with a significant component along the axial $[110]$ direction, which 
we have labelled $(100)_\perp$, irrespective of vector direction. Bonds with a very small vector component 
along the $[110]$ direction along with a significant
vector component perpendicular to the $(111)$ surface in the first quadrant, which we have labelled
$(111)_\perp$. The final bond type, which is symmetrically equivalent to $(111)_\perp$ in the second quadrant,
we have chosen to label $(111)_\parallel$, as it forms the $(111)$ surfaces in the first quadrant.

\subsection{SiGe-NWs}

There are many ways to present the data on bond lengths and local strain; we show two in Fig.~\ref{Si3Ge_core}.  The left-hand side of the figure shows a three-dimensional plot of variations in bond length, which gives full information on the structural variation, but is difficult to interpret, and consistent presentation for different size nanowires is almost impossible.  The right-hand side shows the result of projecting the average bond length, for each of the bond types shown in Fig.~\ref{BondLabel}, onto a grid in a plane perpendicular to the nanowire axis (we note that the $(111)_\perp$ bonds are not shown as they are symmetrically equivalent to the $(111)_{\parallel}$ bonds).  The variation in strain with location in the nanowire and with direction is clearly visible, showing the effects of the surface reconstruction as well as local strains.  A careful inspection will show that some information is lost: for instance, in the third layer below the (100) surfaces, the surface reconstruction induces an alternating compressive and tensile strain \emph{along} the nanowire axis which is not seen in the averaged plots.  However, this is a small variation, and does not affect the overall conclusions.

% SiGe SHELL
\begin{figure}[t]
  \begin{center}
  \begin{tabular}{c c}
    \begin{tabular}{c c c }
    \hline
       NW Model & $(111)_\perp$ & $(100)_\perp$\\
    \hline
       \textbf{3\_3} &
       \begin{minipage}[h]{0.14\textwidth}
          \includegraphics[trim = 30mm 30mm 40mm 30mm, clip, width=1.\textwidth]{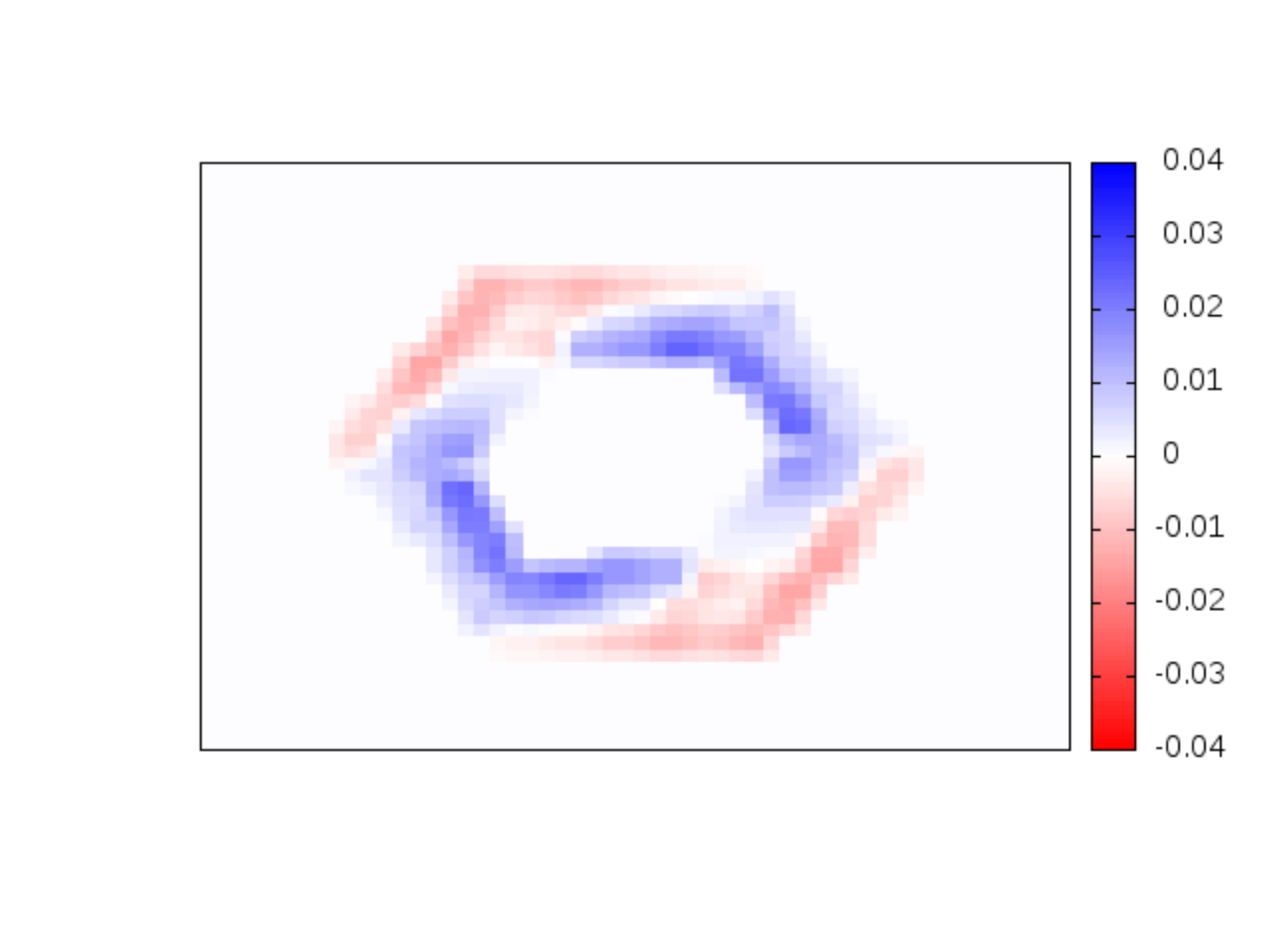}
        \end{minipage}&
       \begin{minipage}[h]{0.14\textwidth}
          \includegraphics[trim = 30mm 30mm 40mm 30mm, clip, width=1.\textwidth]{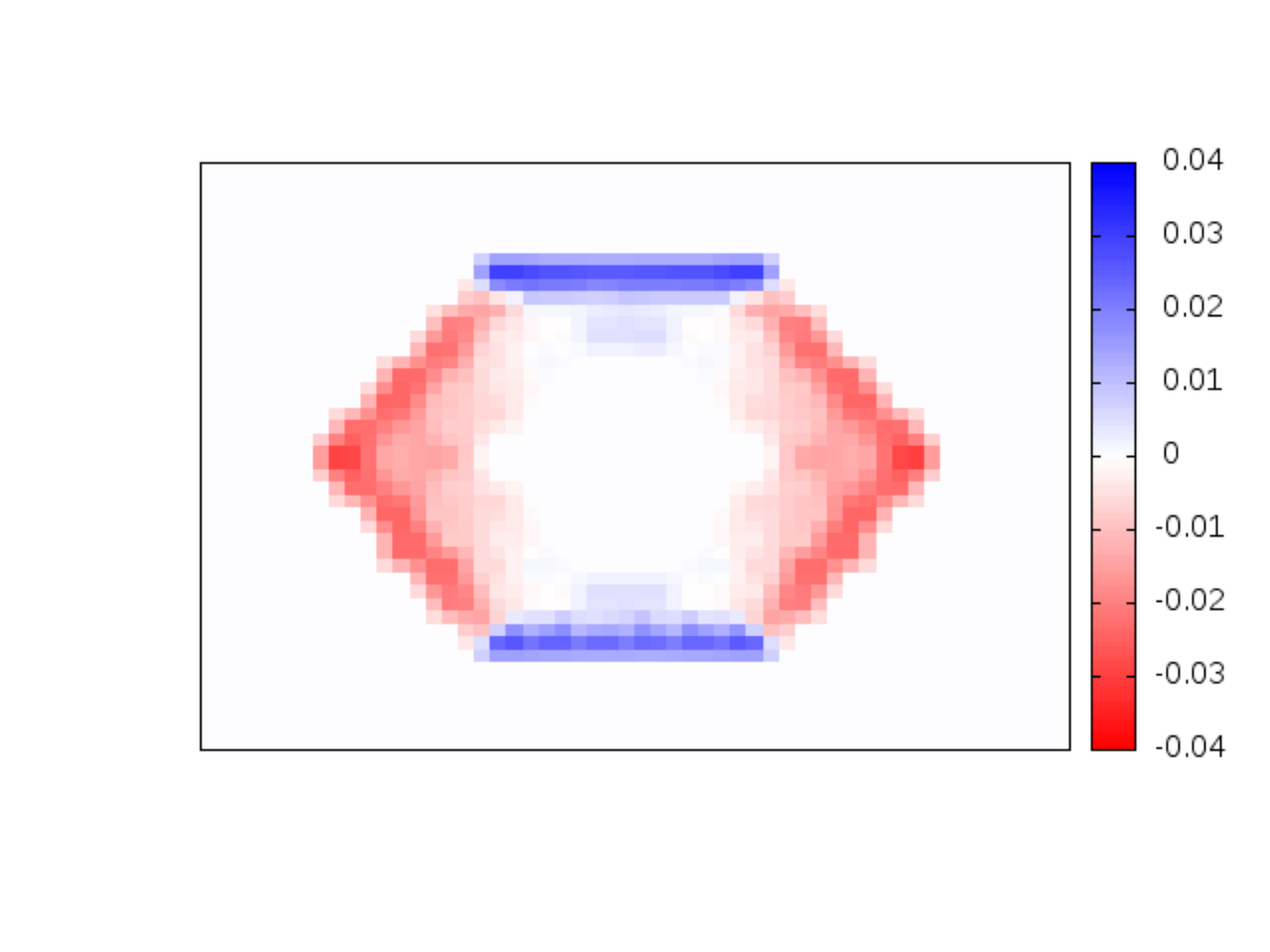}
        \end{minipage}\\
        \textbf{3\_5} &
        \begin{minipage}[h]{0.14\textwidth}
          \includegraphics[trim = 30mm 30mm 40mm 30mm, clip, width=1.\textwidth]{IMAGES/S3G5_shell_map_1.pdf}
        \end{minipage}&
       \begin{minipage}[h]{0.14\textwidth}
          \includegraphics[trim = 30mm 30mm 40mm 30mm, clip, width=1.\textwidth]{IMAGES/S3G5_shell_map_3.pdf}
        \end{minipage}\\
        \textbf{3\_7}&
         \begin{minipage}[h]{0.14\textwidth}
          \includegraphics[trim = 30mm 30mm 40mm 30mm, clip, width=1.\textwidth]{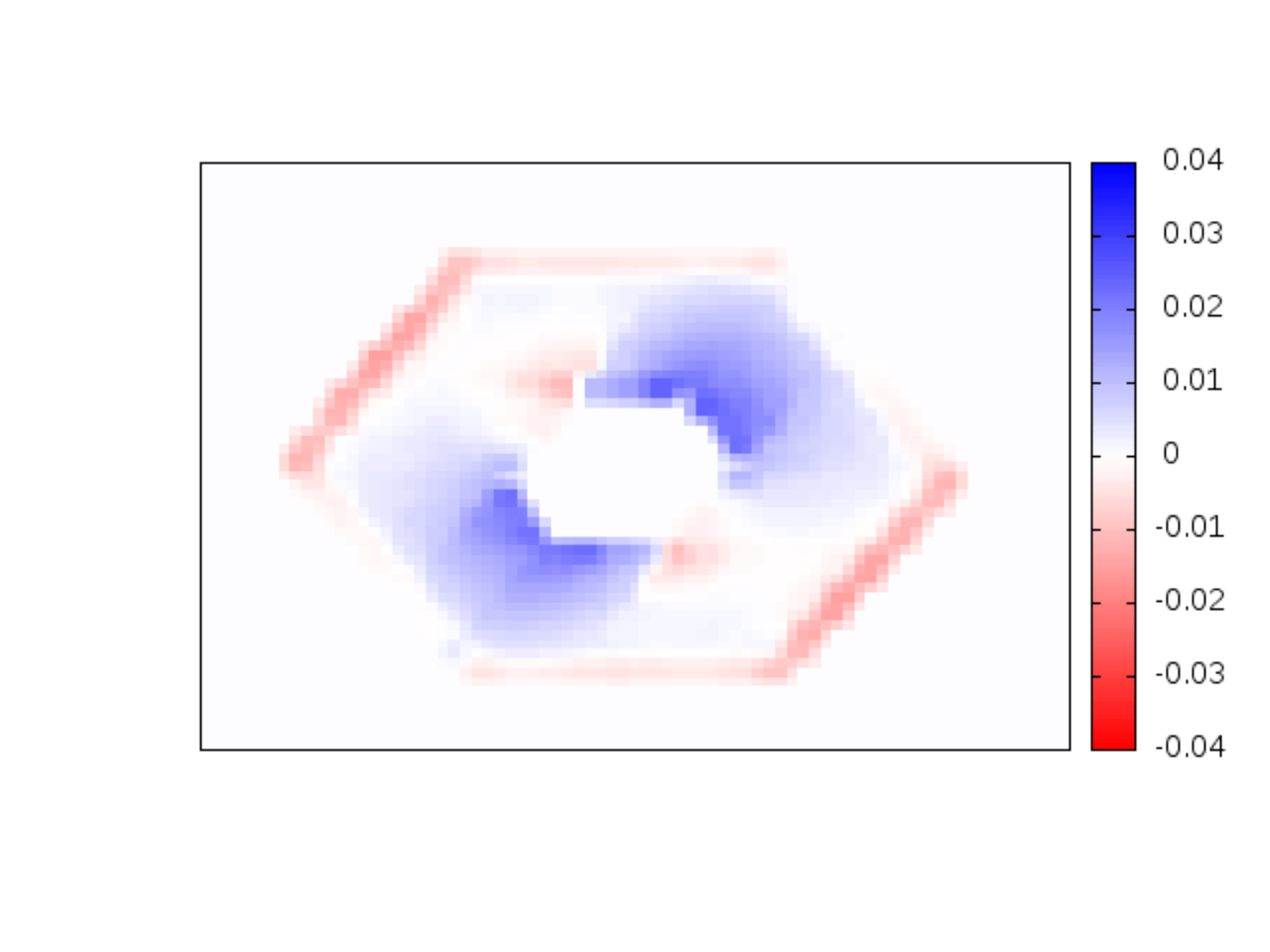}
        \end{minipage}&
       \begin{minipage}[h]{0.14\textwidth}
          \includegraphics[trim = 30mm 30mm 40mm 30mm, clip, width=1.\textwidth]{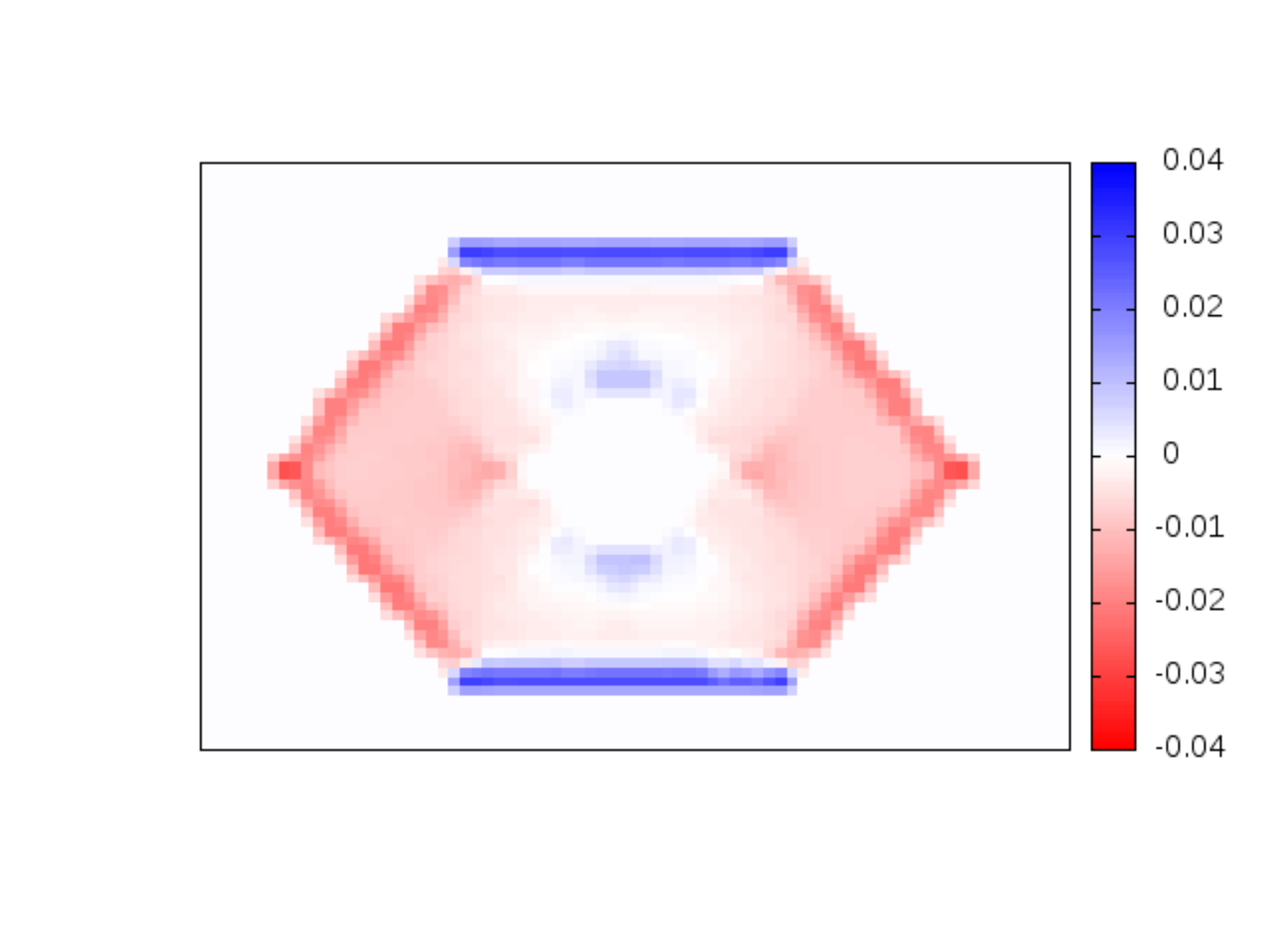}
        \end{minipage}\\
        \textbf{6\_3}&
         \begin{minipage}[h]{0.14\textwidth}
          \includegraphics[trim = 30mm 30mm 40mm 30mm, clip, width=1.\textwidth]{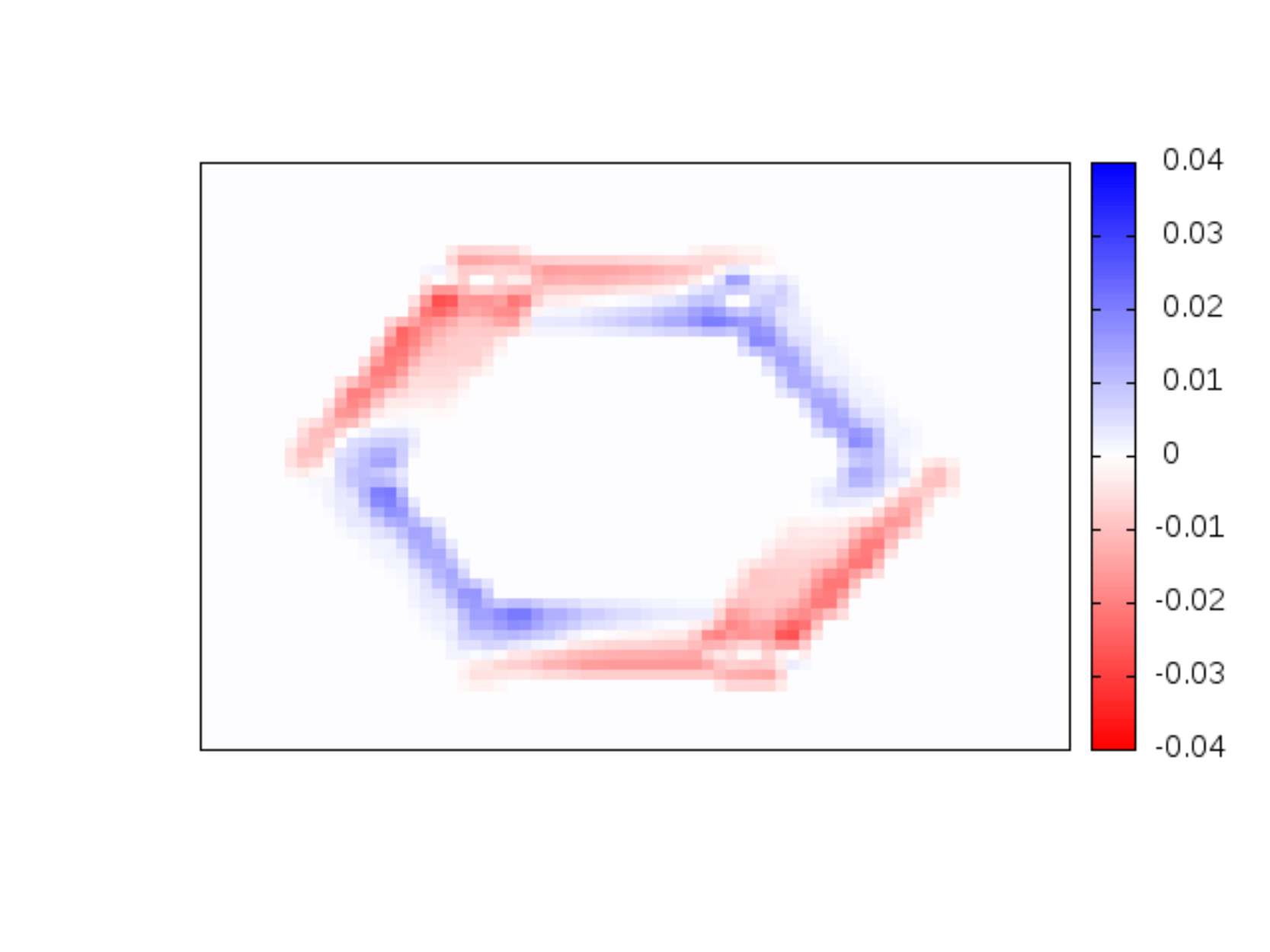}
        \end{minipage}&
       \begin{minipage}[h]{0.14\textwidth}
          \includegraphics[trim = 30mm 30mm 40mm 30mm, clip, width=1.\textwidth]{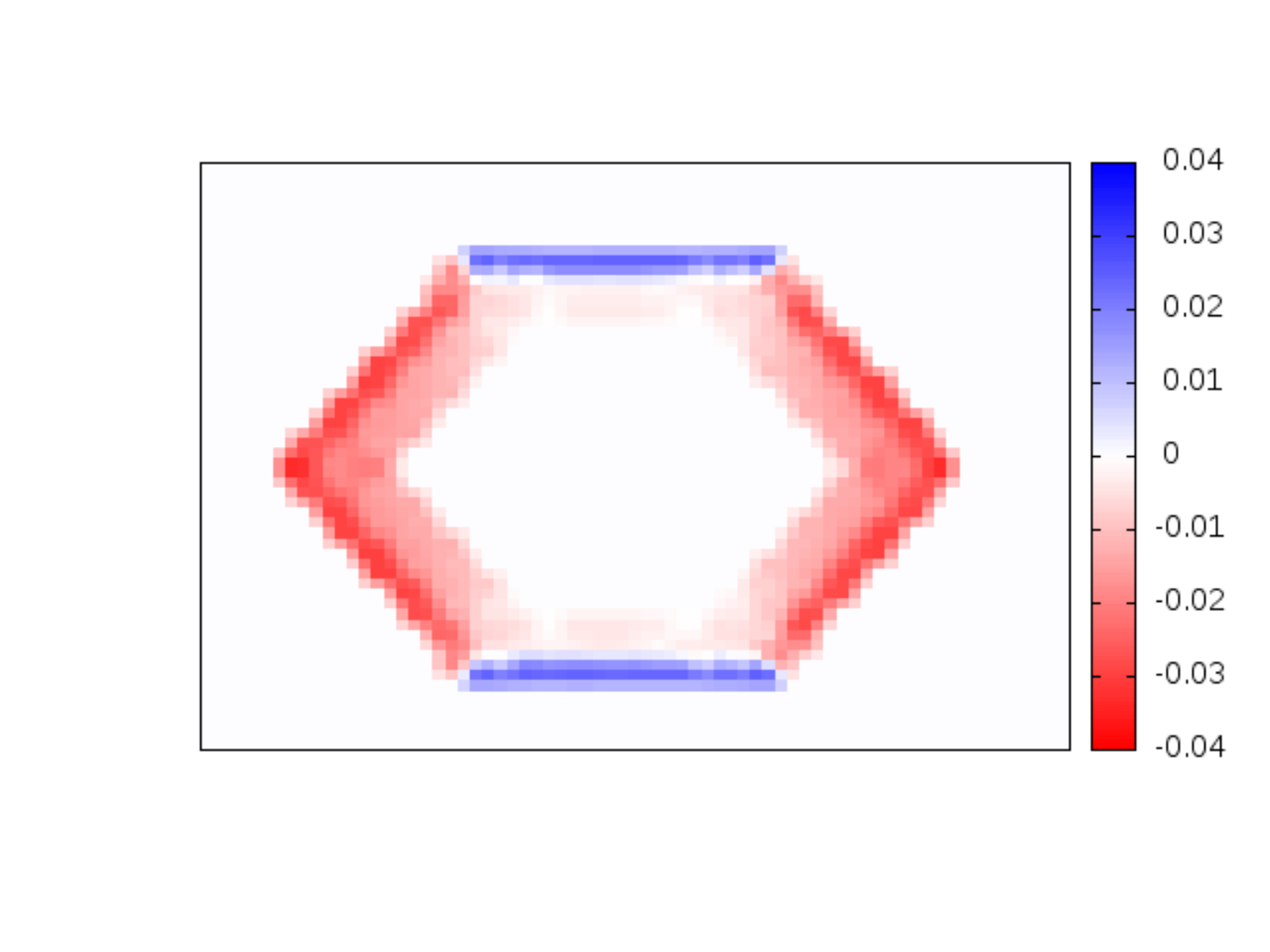}
        \end{minipage}\\
        \textbf{6\_5}&
         \begin{minipage}[h]{0.14\textwidth}
          \includegraphics[trim = 30mm 30mm 40mm 30mm, clip, width=1.\textwidth]{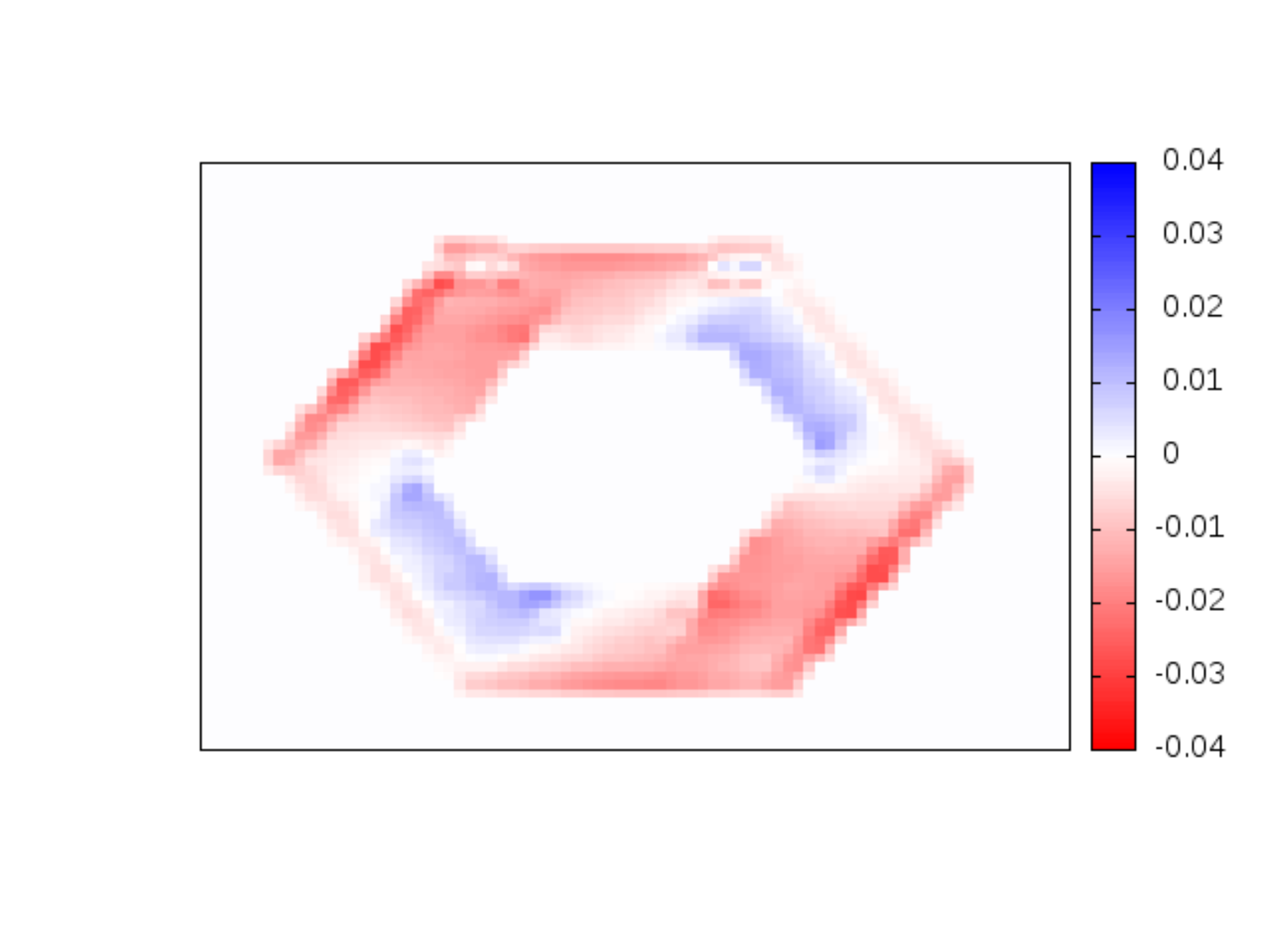}
        \end{minipage}&
       \begin{minipage}[h]{0.14\textwidth}
          \includegraphics[trim = 30mm 30mm 40mm 30mm, clip, width=1.\textwidth]{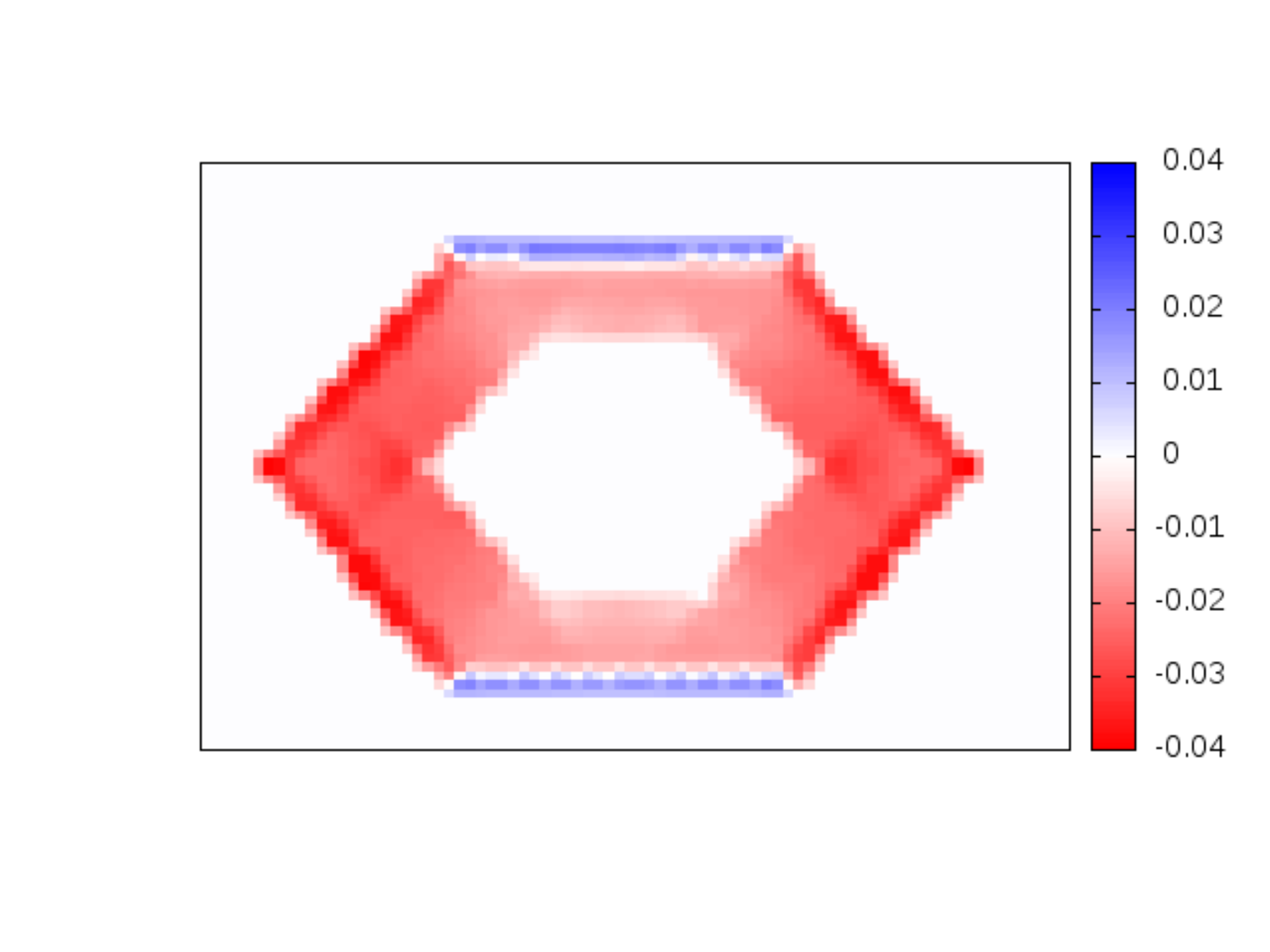}
        \end{minipage}\\
        \textbf{6\_7}&
        \begin{minipage}[h]{0.14\textwidth}
          \includegraphics[trim = 30mm 30mm 40mm 30mm, clip, width=1.\textwidth]{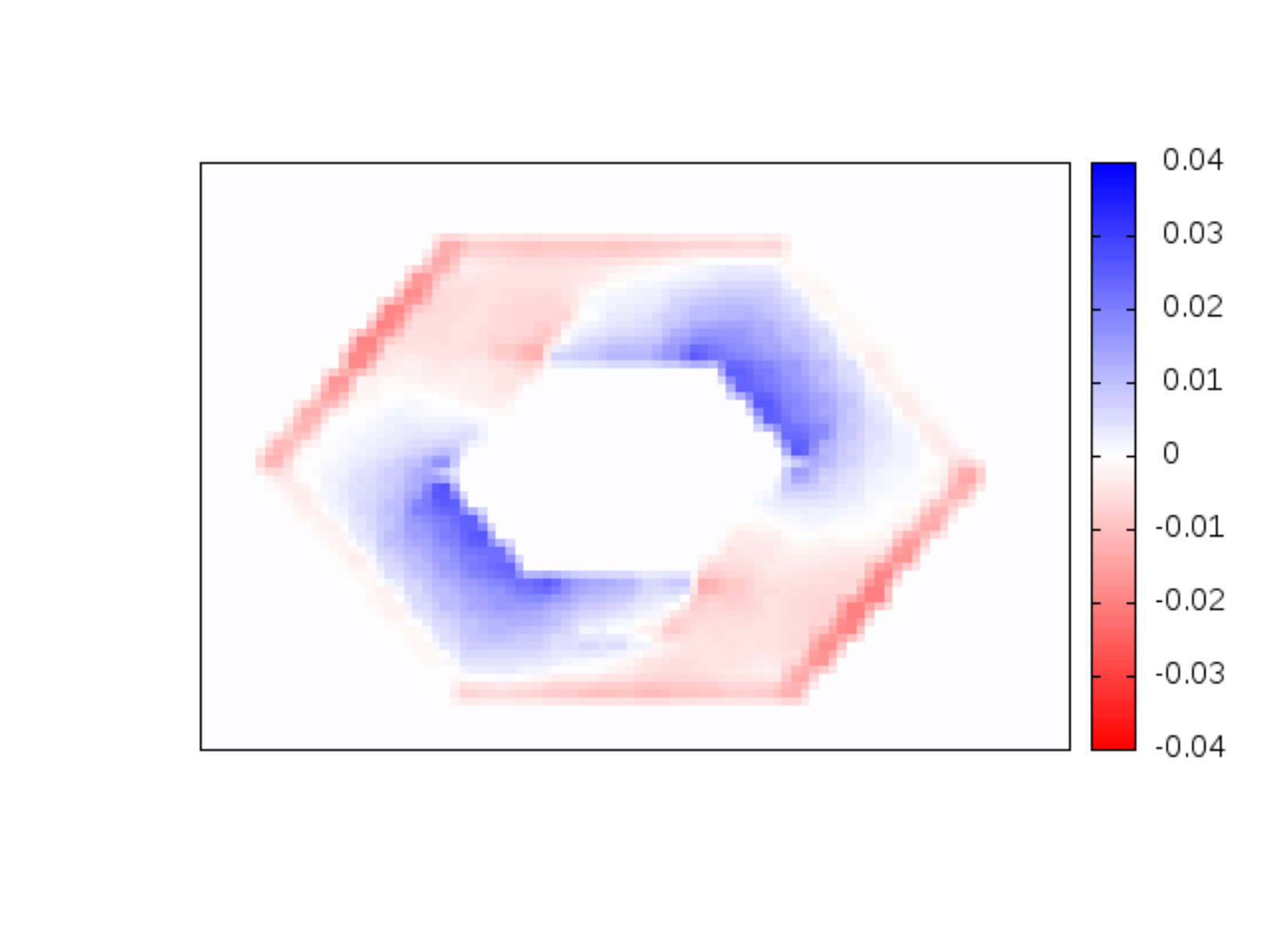}
        \end{minipage}&
       \begin{minipage}[h]{0.14\textwidth}
          \includegraphics[trim = 30mm 30mm 40mm 30mm, clip, width=1.\textwidth]{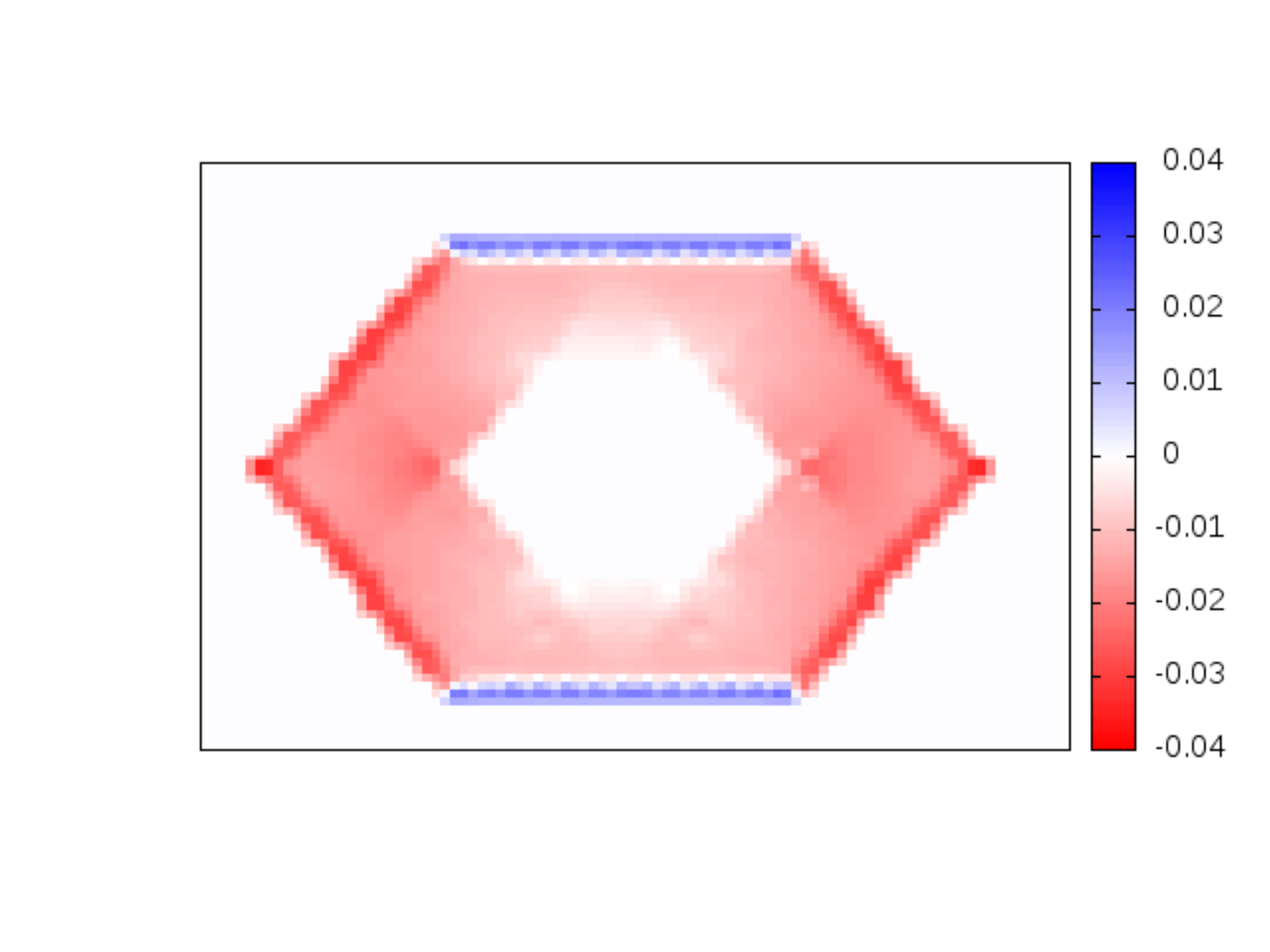}
        \end{minipage}\\
       \hline
    \end{tabular}
    &
    \begin{minipage}[h]{0.065\textwidth}
          \includegraphics[trim = 0mm 0mm 0mm 0mm, clip, width=1.\textwidth]{IMAGES/S3G7_shell_map_label.pdf}
        \end{minipage}\\
    \end{tabular}
    \caption{Average bond strain map for the cross-section of the Ge shell of the SiGe-NWs. Maps for
the $(111)_\perp$ and $(100)_\perp$ labelled bonds are shown, with extension illustrated in blue and compression in red. }
     \label{shell_bonds_SiGe}
  \end{center}
\end{figure}

% SiGe CORE
\begin{figure}[t]
  \begin{center}
  \begin{tabular}{c c}
    \begin{tabular}{c c c }
    \hline
       NW Model & $(111)_\perp$ & $(100)_\perp$\\
    \hline
       \textbf{3\_3} &
       \begin{minipage}[h]{0.14\textwidth}
          \includegraphics[trim = 40mm 40mm 50mm 40mm, clip, width=1.\textwidth]{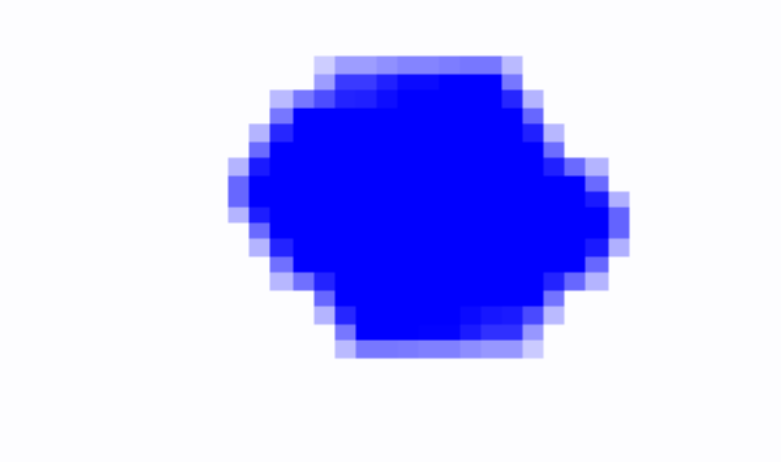}
        \end{minipage}&
       \begin{minipage}[h]{0.14\textwidth}
          \includegraphics[trim = 40mm 40mm 50mm 40mm, clip, width=1.\textwidth]{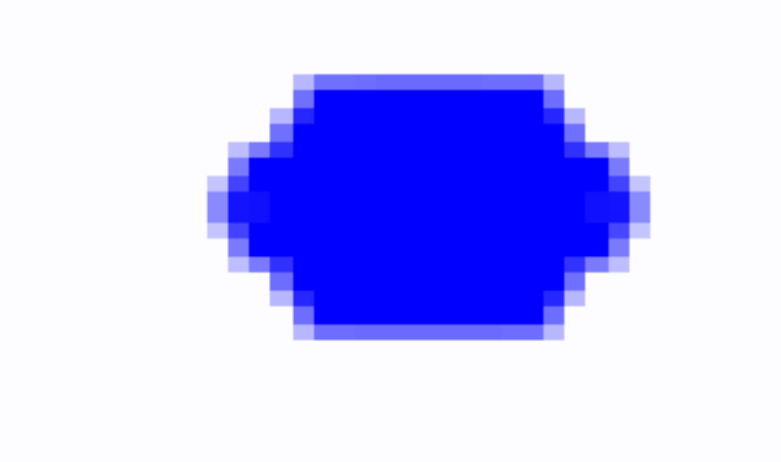}
        \end{minipage}\\
        \textbf{3\_5} &
        \begin{minipage}[h]{0.14\textwidth}
          \includegraphics[trim = 45mm 45mm 55mm 45mm, clip, width=1.\textwidth]{IMAGES/S3G5_core_map_1.pdf}
        \end{minipage}&
       \begin{minipage}[h]{0.14\textwidth}
          \includegraphics[trim = 45mm 45mm 55mm 45mm, clip, width=1.\textwidth]{IMAGES/S3G5_core_map_3.pdf}
        \end{minipage}\\
        \textbf{3\_7}&
         \begin{minipage}[h]{0.14\textwidth}
          \includegraphics[trim = 50mm 50mm 60mm 50mm, clip, width=1.\textwidth]{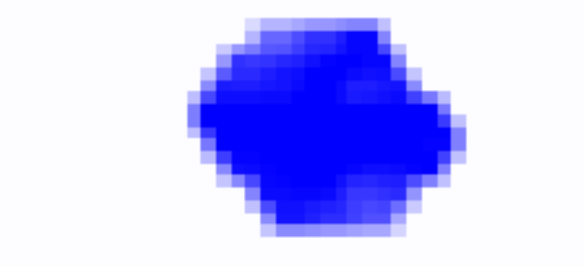}
        \end{minipage}&
       \begin{minipage}[h]{0.14\textwidth}
          \includegraphics[trim = 50mm 50mm 60mm 50mm, clip, width=1.\textwidth]{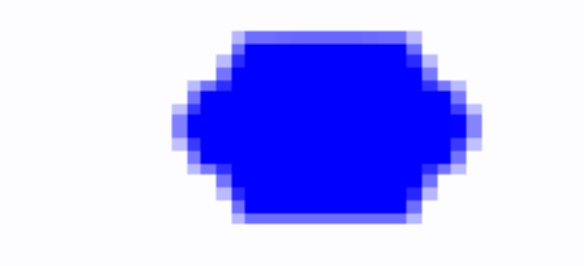}
        \end{minipage}\\
        \textbf{6\_3}&
         \begin{minipage}[h]{0.14\textwidth}
          \includegraphics[trim = 30mm 30mm 40mm 30mm, clip, width=1.\textwidth]{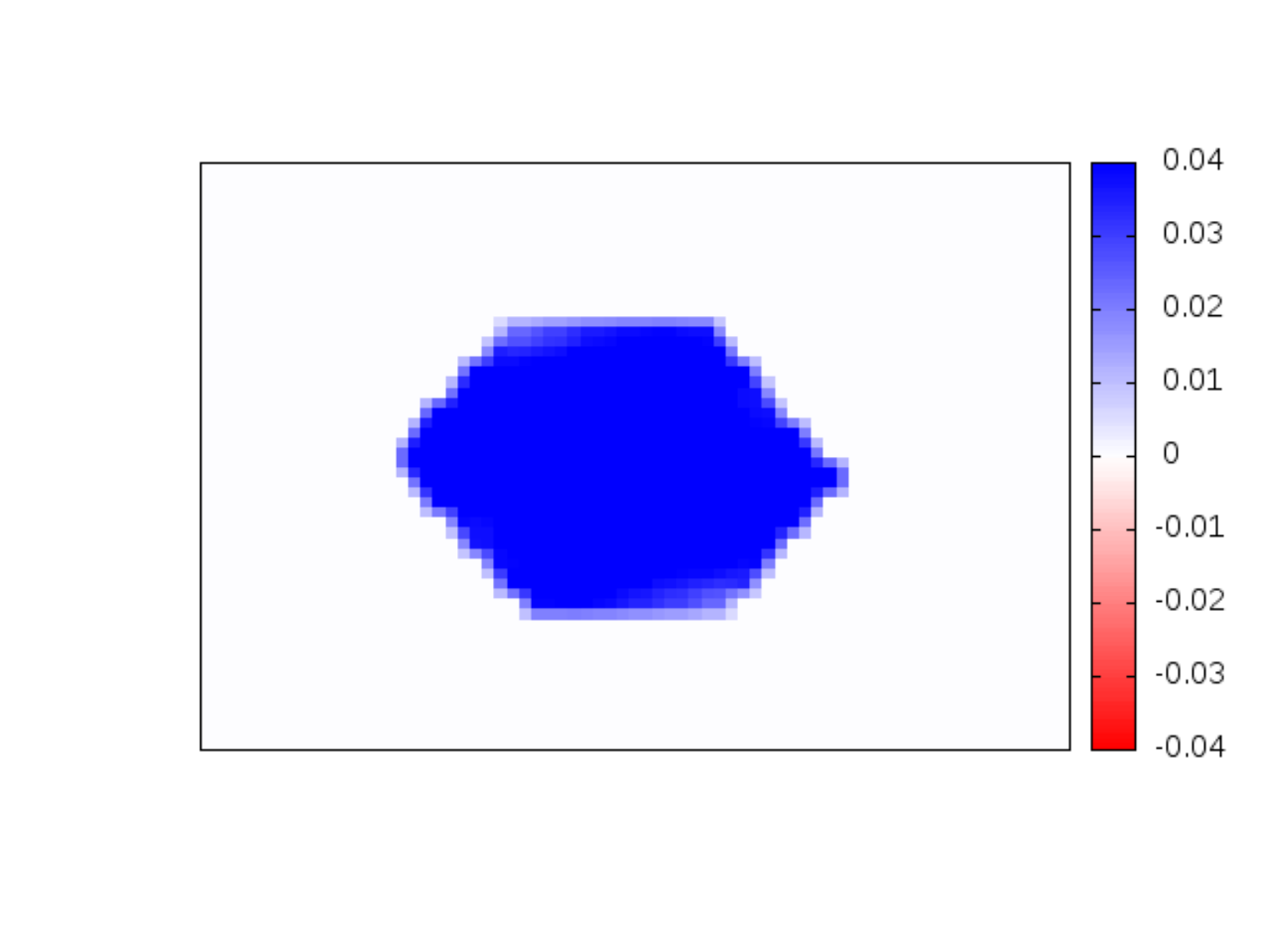}
        \end{minipage}&
       \begin{minipage}[h]{0.14\textwidth}
          \includegraphics[trim = 30mm 30mm 40mm 30mm, clip, width=1.\textwidth]{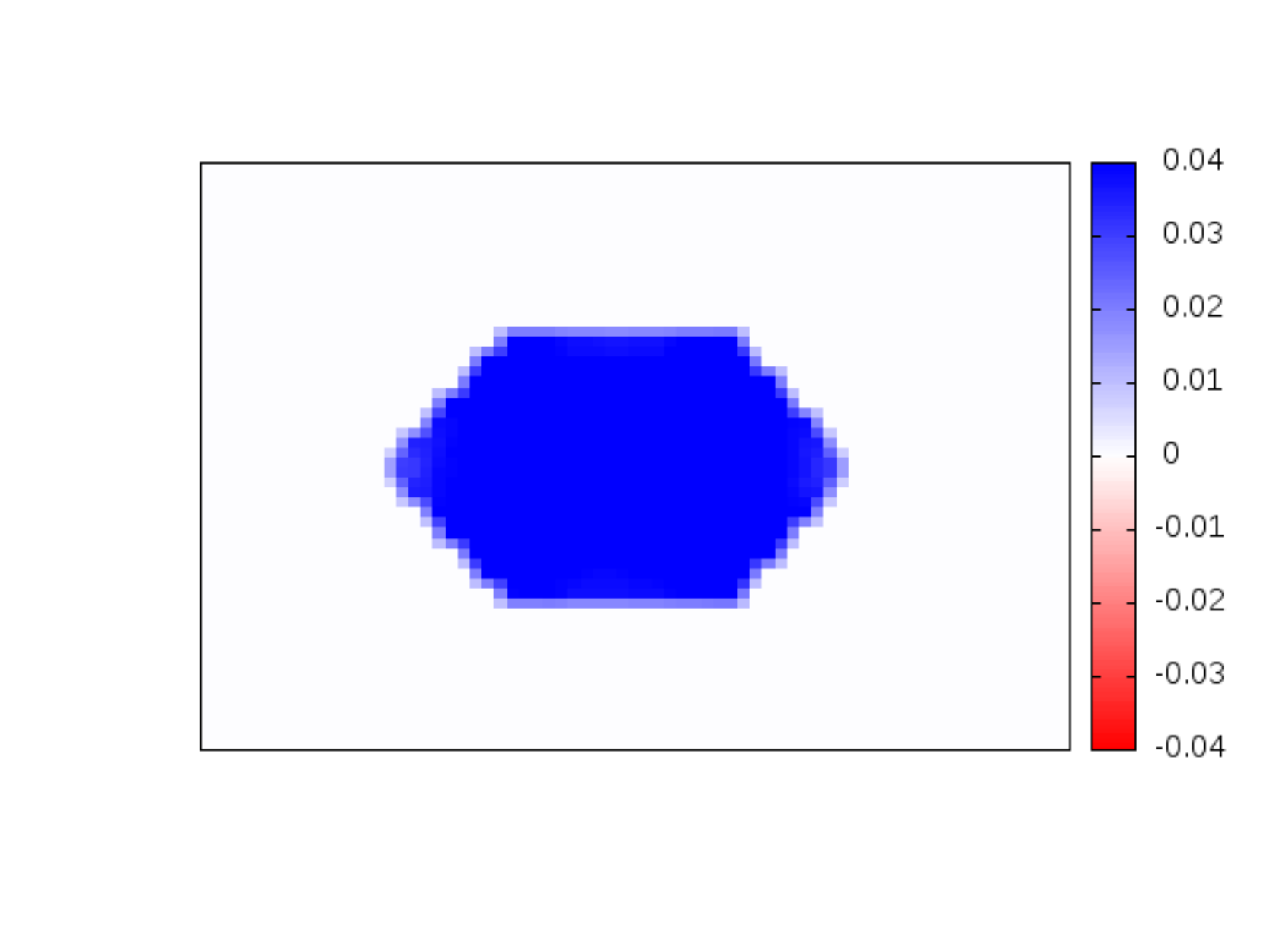}
        \end{minipage}\\
        \textbf{6\_5}&
         \begin{minipage}[h]{0.14\textwidth}
          \includegraphics[trim = 35mm 35mm 45mm 35mm, clip, width=1.\textwidth]{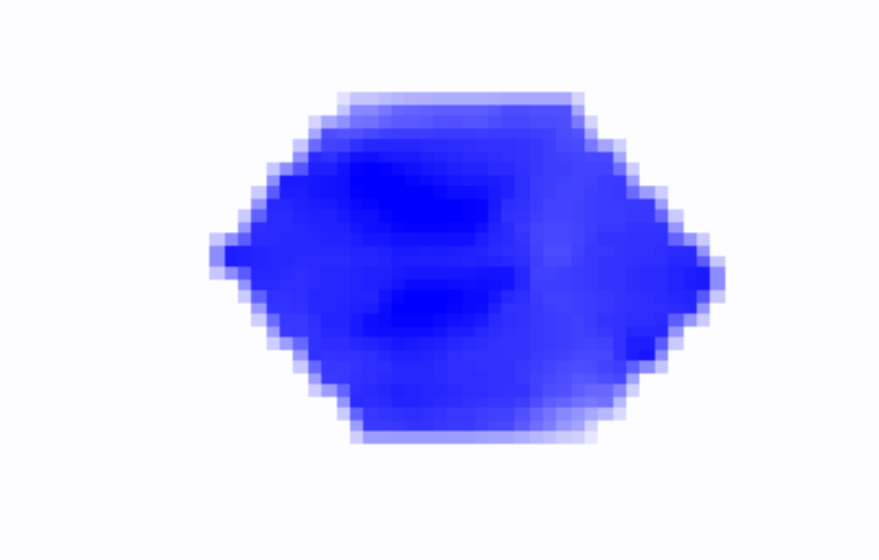}
        \end{minipage}&
       \begin{minipage}[h]{0.14\textwidth}
          \includegraphics[trim = 35mm 35mm 45mm 35mm, clip, width=1.\textwidth]{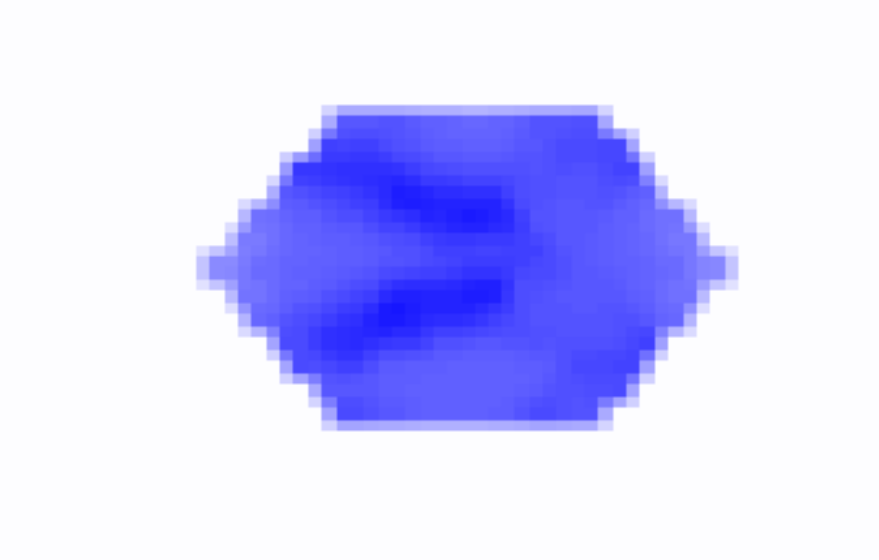}
        \end{minipage}\\
        \textbf{6\_7}&
        \begin{minipage}[h]{0.14\textwidth}
          \includegraphics[trim = 40mm 40mm 50mm 40mm, clip, width=1.\textwidth]{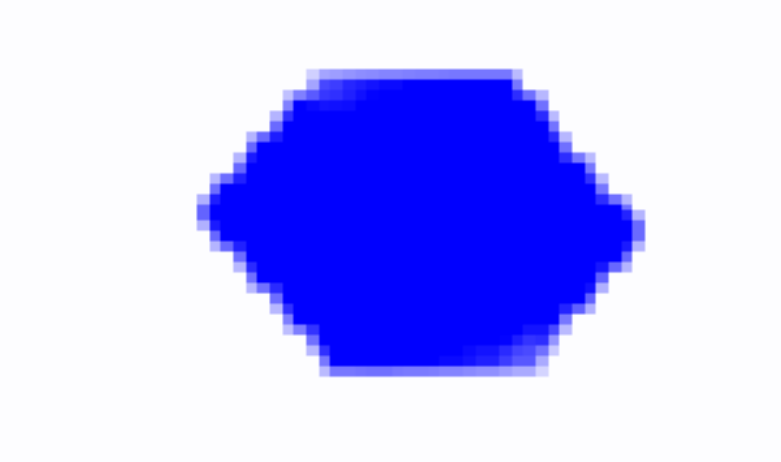}
        \end{minipage}&
       \begin{minipage}[h]{0.14\textwidth}
          \includegraphics[trim = 40mm 40mm 50mm 40mm, clip, width=1.\textwidth]{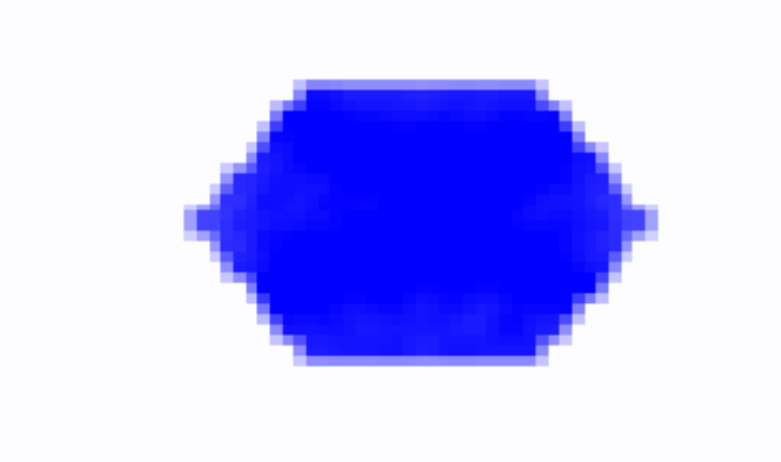}
        \end{minipage}\\
       \hline
    \end{tabular}
    &
    \begin{minipage}[h]{0.065\textwidth}
          \includegraphics[trim = 0mm 0mm 0mm 0mm, clip, width=1.\textwidth]{IMAGES/S3G7_shell_map_label.pdf}
        \end{minipage}\\
    \end{tabular}
    \caption{Average bond strain map for the cross-section of the Si core of the SiGe-NWs. Maps for
the $(111)_\perp$ and $(100)_\perp$ labelled bonds are shown, with extension illustrated in blue and compression in red. }
     \label{core_bonds_SiGe}
  \end{center}
\end{figure}

The bond strain maps for the Si core, Ge shell nanowires are shown in Fig.~\ref{shell_bonds_SiGe} (shell) and Fig.~\ref{core_bonds_SiGe} (core).  
Significant anisotropy within the shell becomes apparent, with bonds in the $(111) _\perp$ directions
behaving very differently to the $(100)_\perp$ bonds. 
These bonds are extended when the bond direction is perpendicular to the closest $(111)$ nanowire surface.
This extension is most pronounced close to the Si-Ge interface and monotonically reduces to the bulk 
bond length at the surface. On the other hand we see a slight compression at the Si-Ge interface
in this bond type when it runs parallel to the closest $(111)$ surface, and compression of this
bond in each of the four $(111)$ shell surfaces it forms in each of the four quadrants. The $(100)_\perp$ bond behaves in a significantly different manner, with compression on all of the $(111)$ surfaces and throughout
most of the shell interior. The interior compression is most pronounced at the SiGe interface at the 
point where the two $(111)$ surfaces intersect. However there is also a slight extension 
of this bond type at the $(100)$ Si-Ge interface, as well as a considerable extension due to the $(2\times1)$ reconstruction of $(100)$ Ge surfaces due to Si-Si dimerisation. Overall we can see that the largest strains are at the Si-Ge interfaces and the Ge surfaces.

Turning to the core, shown in Fig.~\ref{core_bonds_SiGe}, we see that it is expanded for all shell thicknesses. Both the 3-layer and 6-layer cores expand, and the expansion is largely isotropic, particularly in comparison to the shell.  This is in agreement with previous theoretical and experimental work\cite{Fukata:2012xm,Peng:2010gk}, in which the Si core is found to be under tensile strain.

It is well known that both Si and Ge are mechanically anisotropic, with the Young's modulus in 
the $[111]$ direction larger than that in the $[110]$, which in turn is larger than that in the $[100]$
direction. It is clear that this anisotropy has a significant impact upon the relative bond compression 
and extension in each direction within the Ge shell. Coupling this fact with the surface and interface 
effects result in the differing strain patterns of the bonds depending upon bond direction. It is interesting to note that the core of the NW does not have the same radial freedom as the shell, which can expand into the vacuum, nor the strains induced by reconstruction; however, as we will see in the next section, for these nanowires and the present method, the anisotropy is largely associated with germanium, while silicon is much more uniformly strained.  (We note that silicon does have a larger Young's modulus than germanium\cite{Wortman:1965yr}, but it is unlikely that the cause is anything this simple.)

\subsection{GeSi-NWs}

% GeSi SHELL
\begin{figure}[t]
  \begin{center}
  \begin{tabular}{c c}
    \begin{tabular}{c c c }
    \hline
       NW Model & $(111)_\perp$ & $(100)_\perp$\\
    \hline
       \textbf{3\_3} &
       \begin{minipage}[h]{0.14\textwidth}
          \includegraphics[trim = 30mm 30mm 40mm 30mm, clip, width=1.\textwidth]{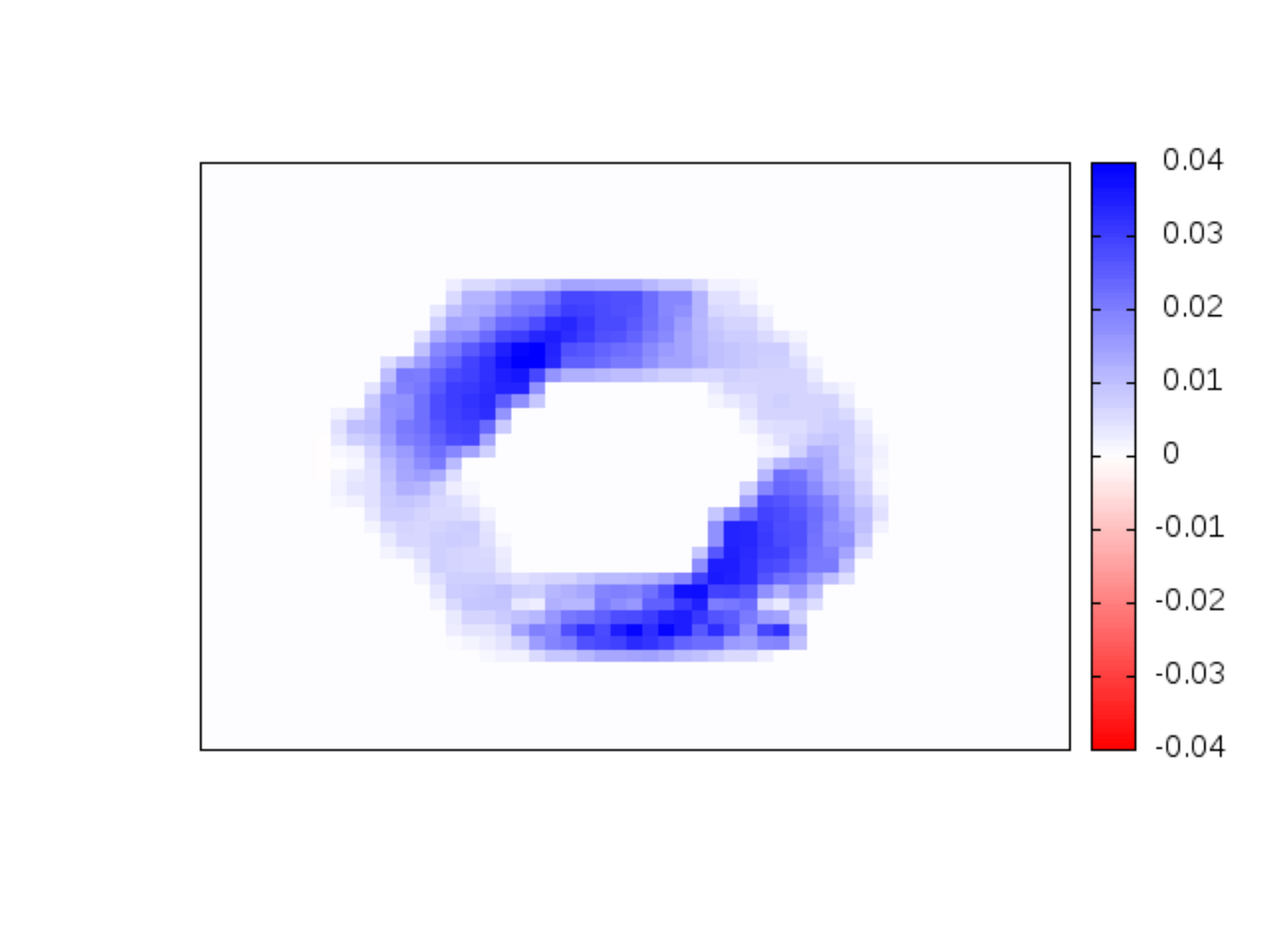}
        \end{minipage}&
       \begin{minipage}[h]{0.14\textwidth}
          \includegraphics[trim = 30mm 30mm 40mm 30mm, clip, width=1.\textwidth]{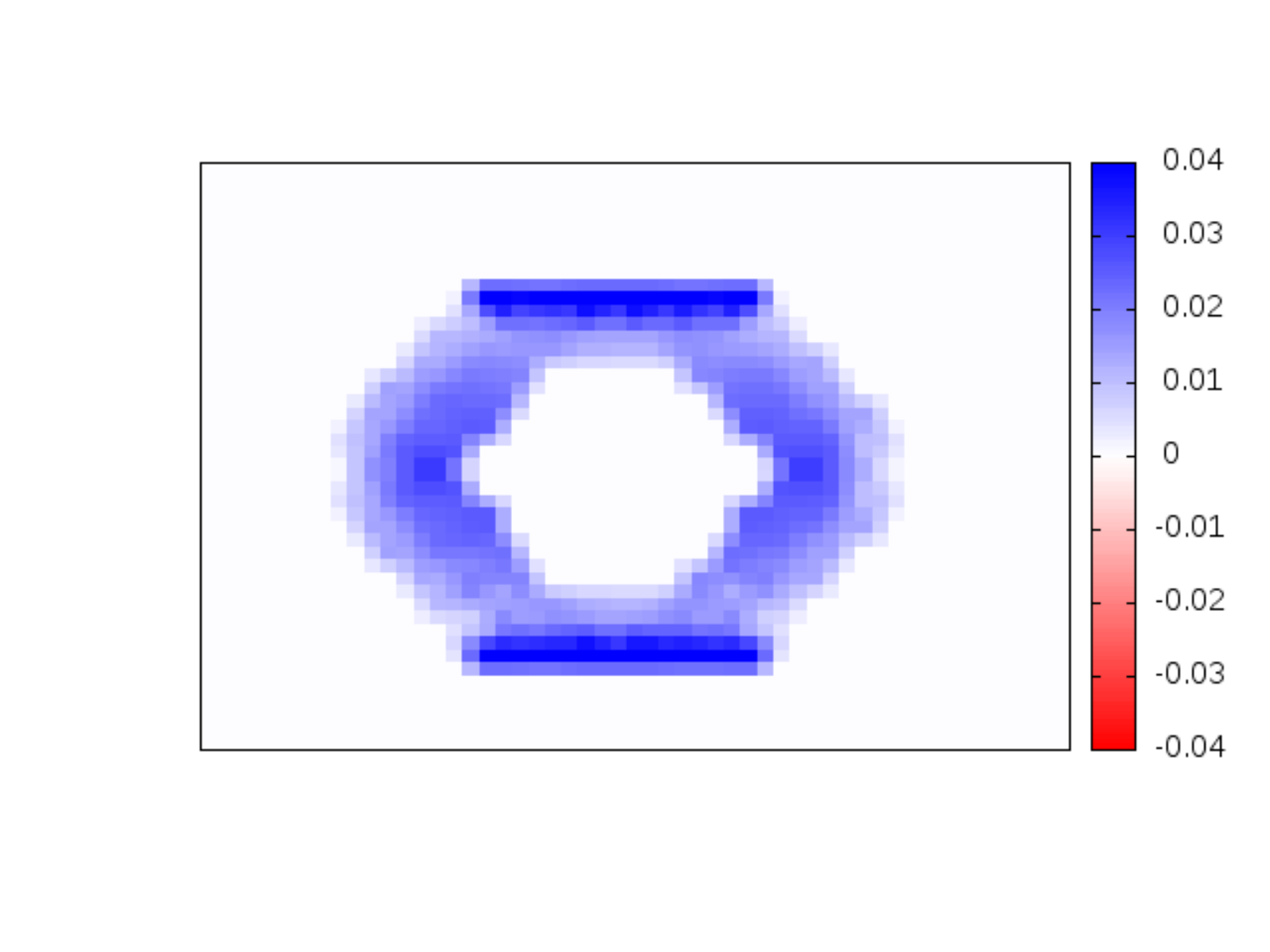}
        \end{minipage}\\
        \textbf{3\_5} &
        \begin{minipage}[h]{0.14\textwidth}
          \includegraphics[trim = 30mm 30mm 40mm 30mm, clip, width=1.\textwidth]{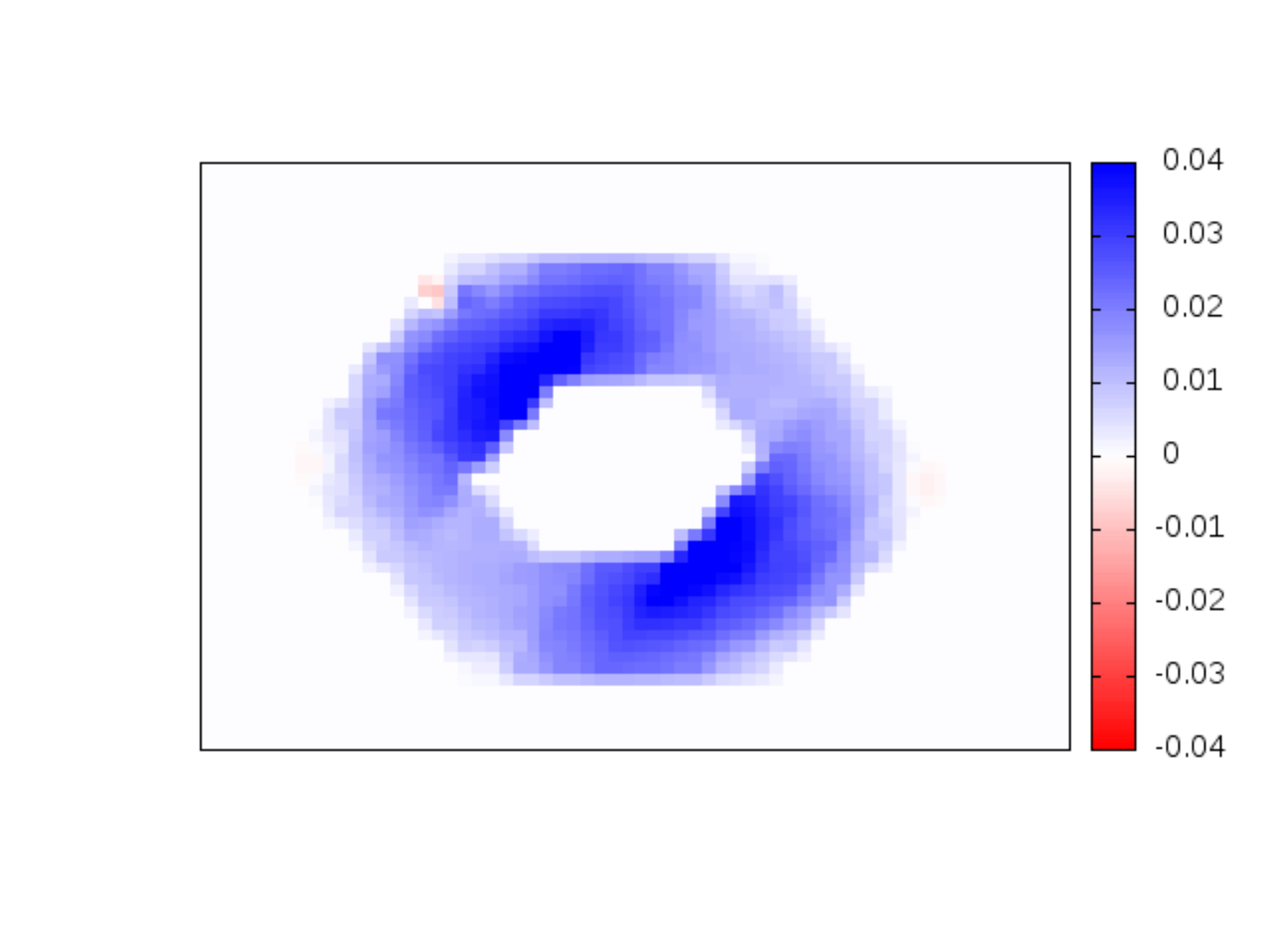}
        \end{minipage}&
       \begin{minipage}[h]{0.14\textwidth}
          \includegraphics[trim = 30mm 30mm 40mm 30mm, clip, width=1.\textwidth]{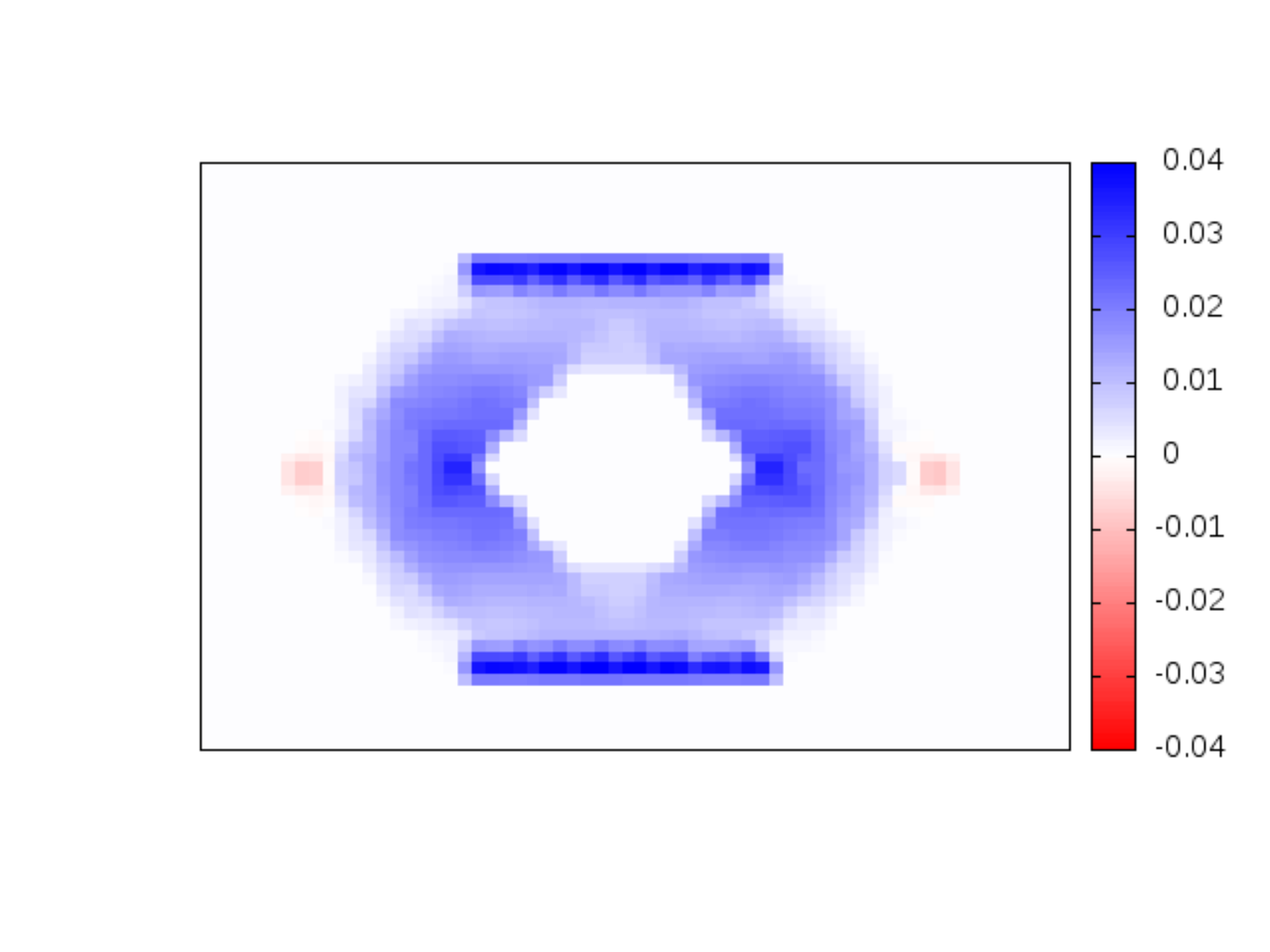}
        \end{minipage}\\
        \textbf{3\_7}&
         \begin{minipage}[h]{0.14\textwidth}
          \includegraphics[trim = 30mm 30mm 40mm 30mm, clip, width=1.\textwidth]{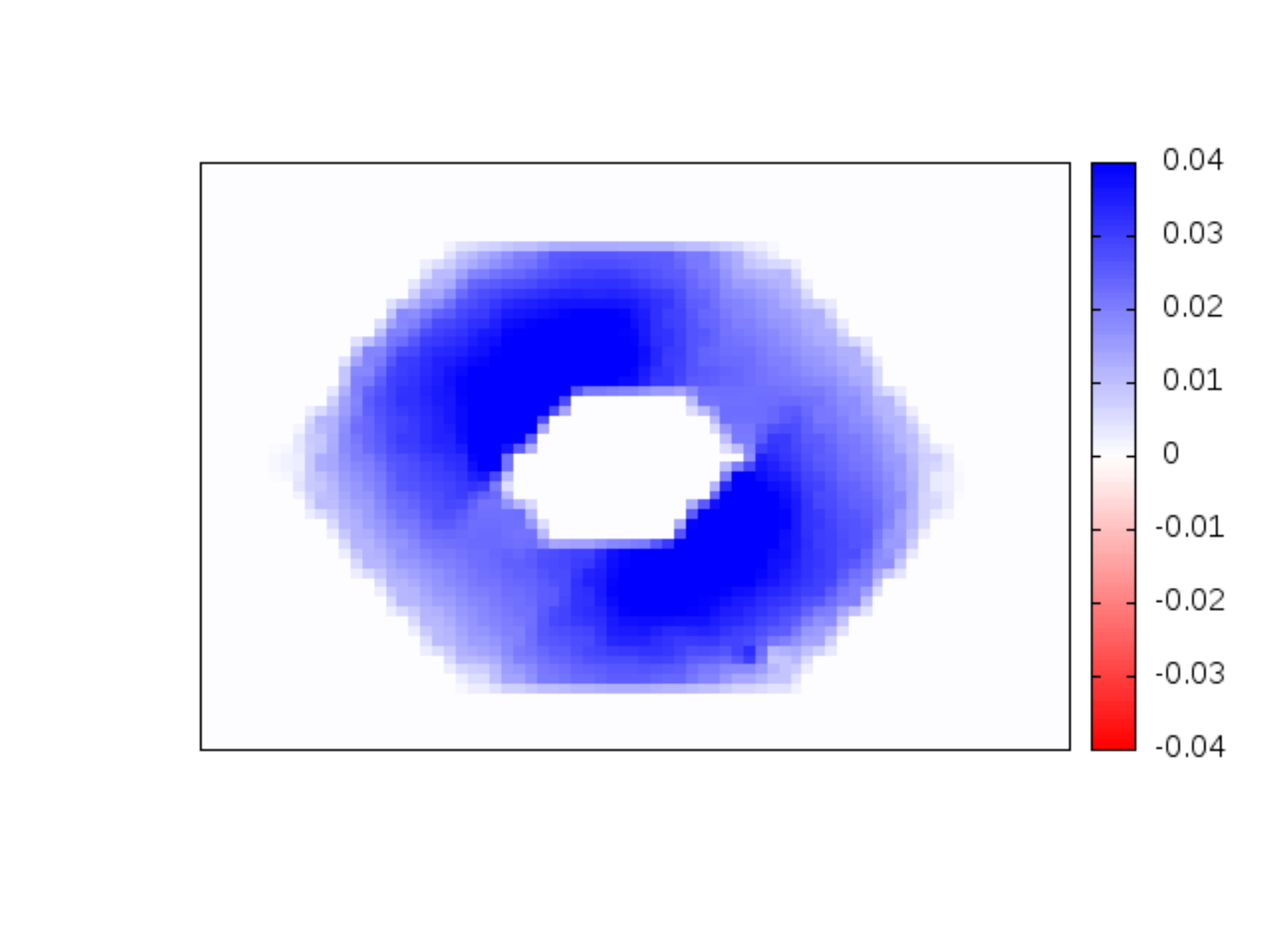}
        \end{minipage}&
       \begin{minipage}[h]{0.14\textwidth}
          \includegraphics[trim = 30mm 30mm 40mm 30mm, clip, width=1.\textwidth]{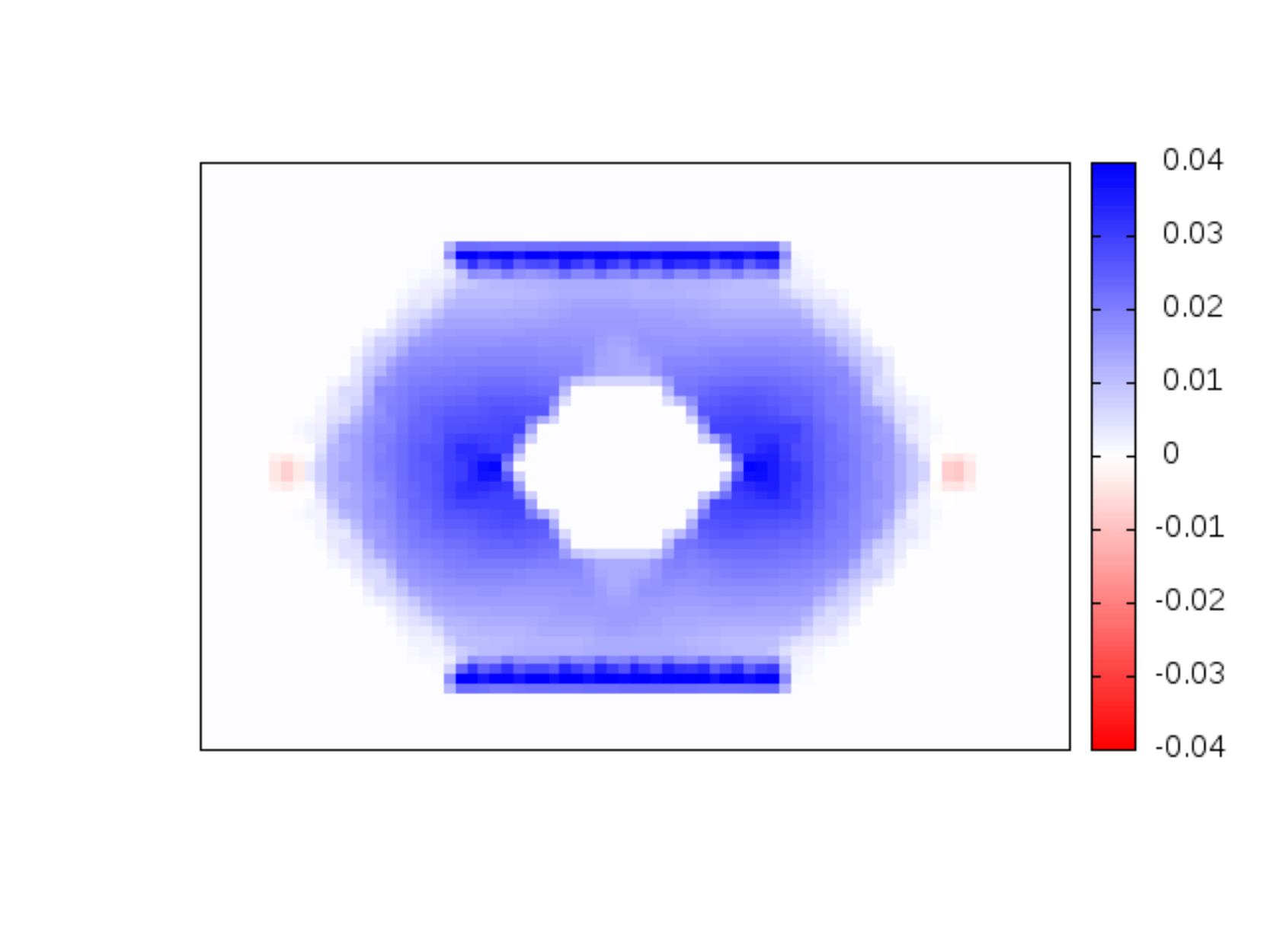}
        \end{minipage}\\
        \textbf{6\_3}&
         \begin{minipage}[h]{0.14\textwidth}
          \includegraphics[trim = 30mm 30mm 40mm 30mm, clip, width=1.\textwidth]{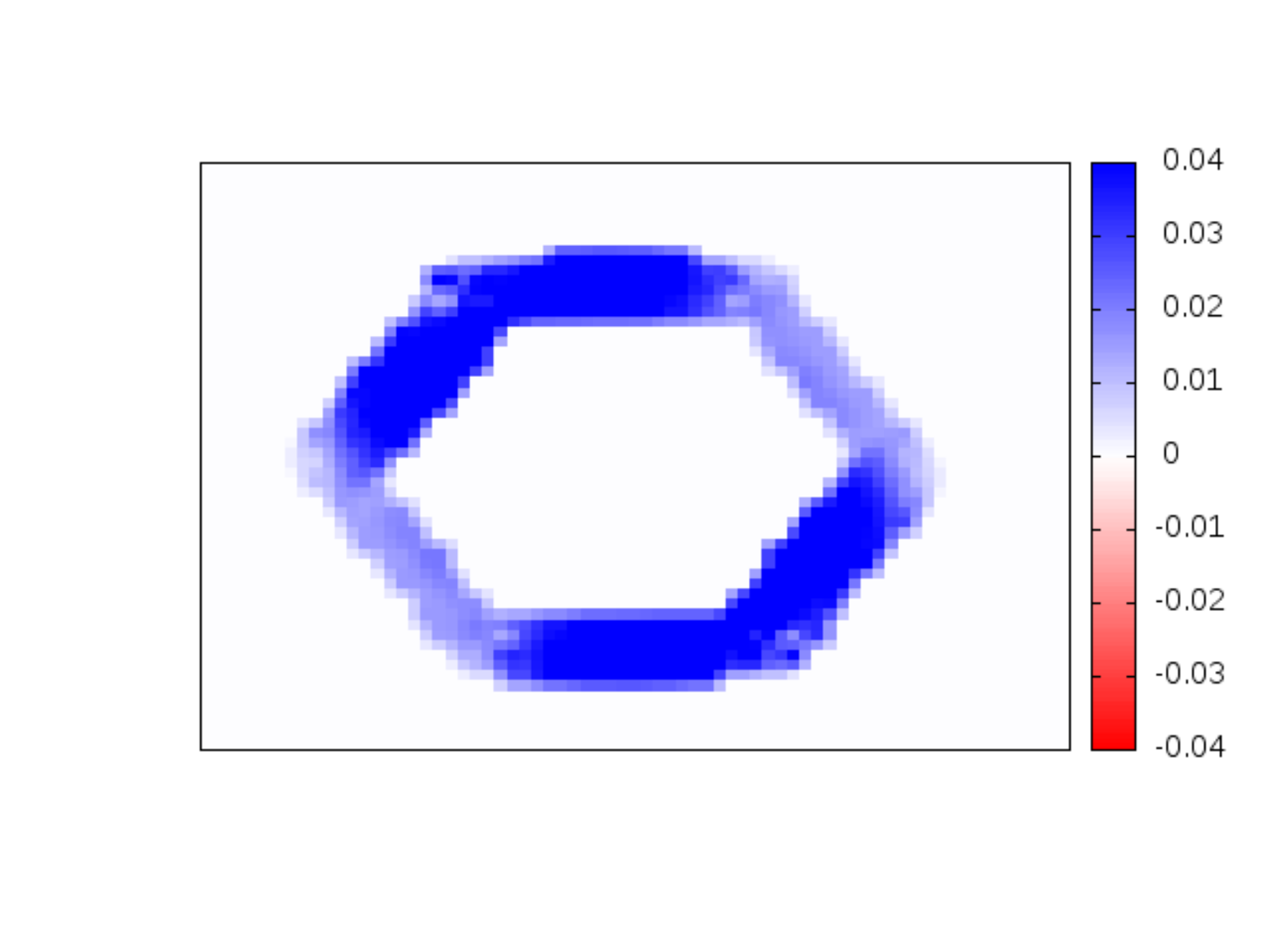}
        \end{minipage}&
       \begin{minipage}[h]{0.14\textwidth}
          \includegraphics[trim = 30mm 30mm 40mm 30mm, clip, width=1.\textwidth]{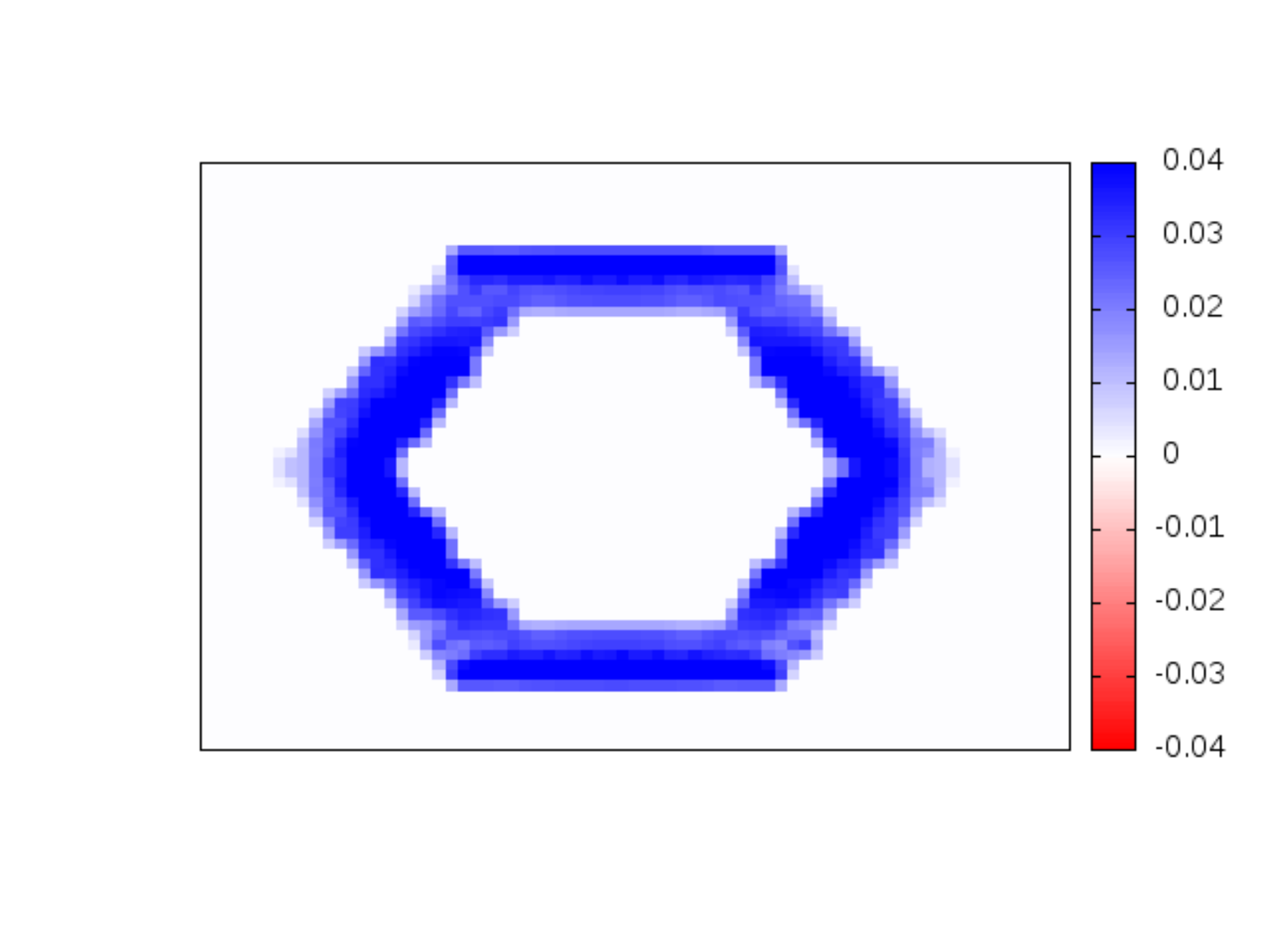}
        \end{minipage}\\
        \textbf{6\_5}&
         \begin{minipage}[h]{0.14\textwidth}
          \includegraphics[trim = 30mm 30mm 40mm 30mm, clip, width=1.\textwidth]{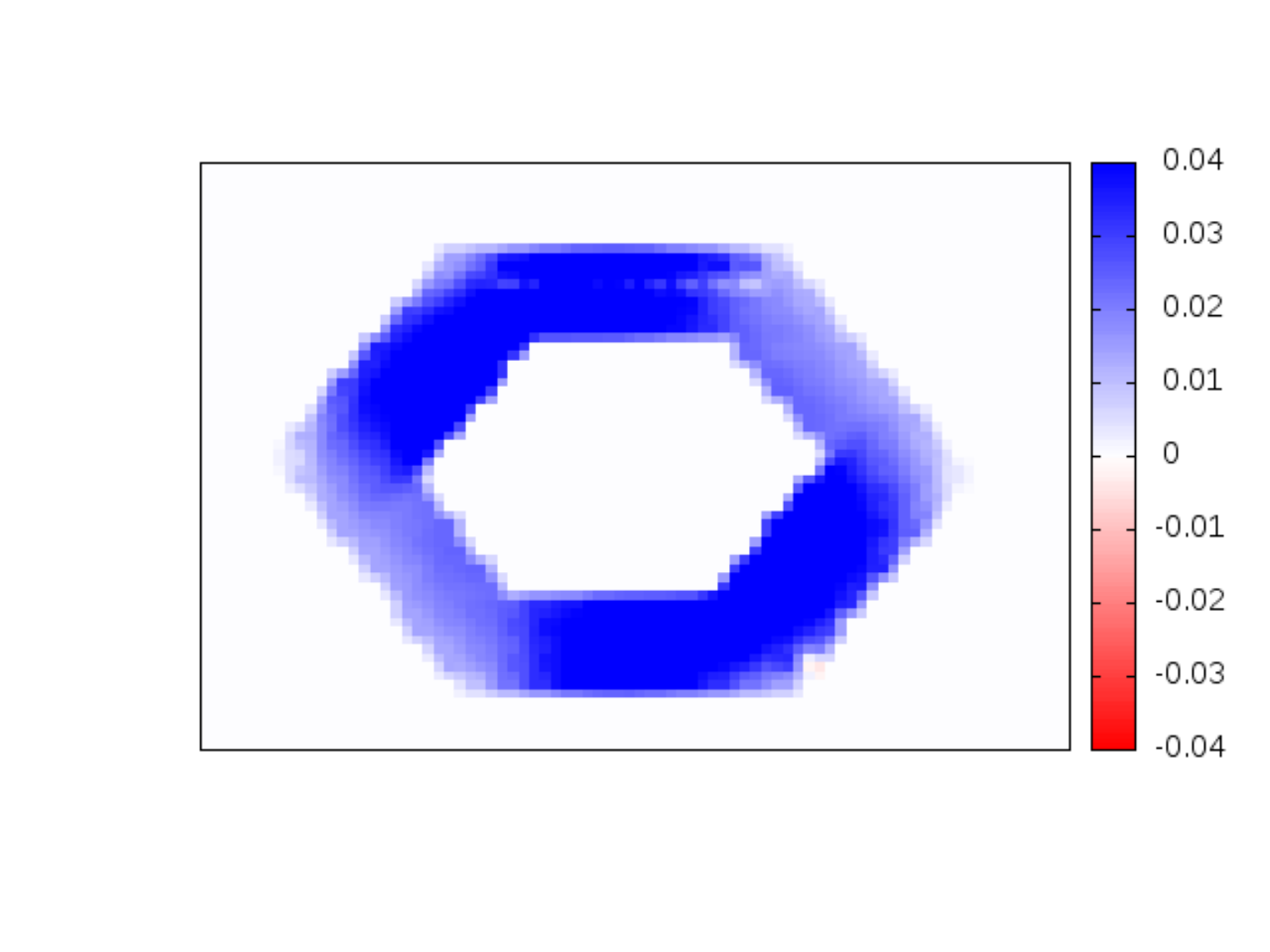}
        \end{minipage}&
       \begin{minipage}[h]{0.14\textwidth}
          \includegraphics[trim = 30mm 30mm 40mm 30mm, clip, width=1.\textwidth]{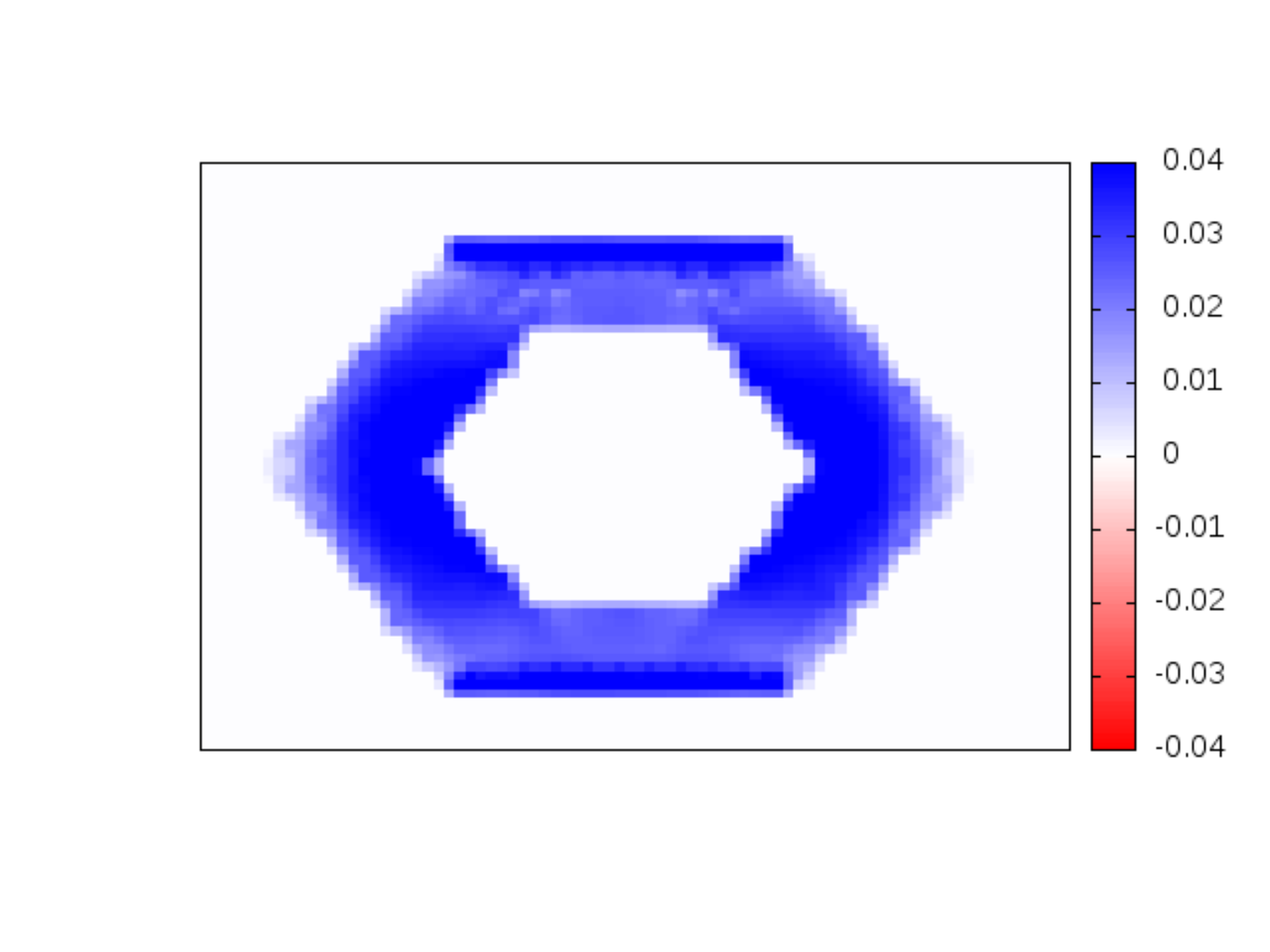}
        \end{minipage}\\
        \textbf{6\_7}&
        \begin{minipage}[h]{0.14\textwidth}
          \includegraphics[trim = 30mm 30mm 40mm 30mm, clip, width=1.\textwidth]{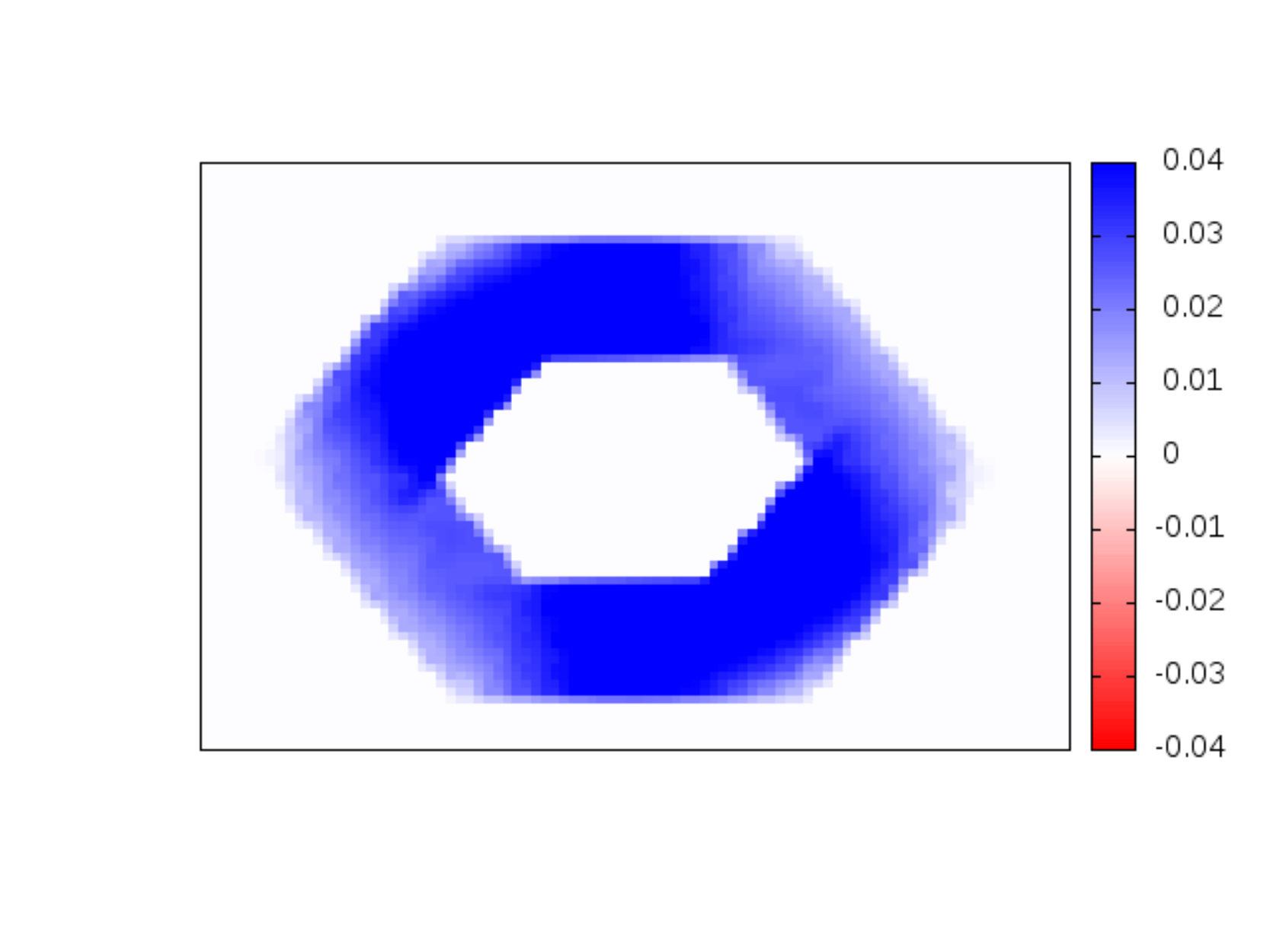}
        \end{minipage}&
       \begin{minipage}[h]{0.14\textwidth}
          \includegraphics[trim = 30mm 30mm 40mm 30mm, clip, width=1.\textwidth]{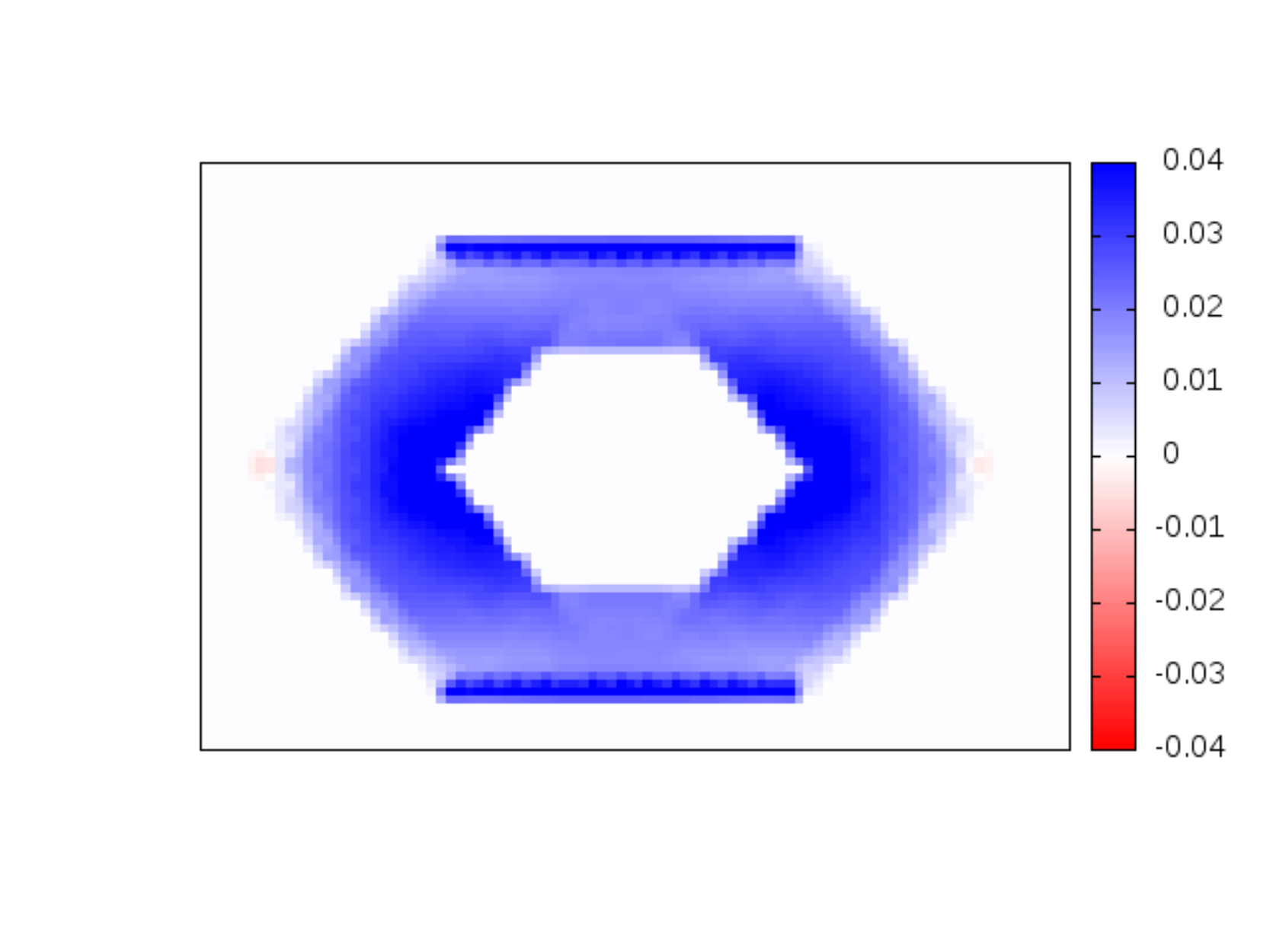}
        \end{minipage}\\
       \hline
    \end{tabular}
    &
    \begin{minipage}[h]{0.065\textwidth}
          \includegraphics[trim = 0mm 0mm 0mm 0mm, clip, width=1.\textwidth]{IMAGES/S3G7_shell_map_label.pdf}
        \end{minipage}\\
    \end{tabular}
    \caption{Average bond strain map for the cross-section of the Si shell of the GeSi-NWs. Maps are shown for the $(111)_\perp$ and $(100)_\perp$ labelled bonds,  with extension illustrated in blue and compression in red. }
     \label{shell_bonds_GeSi}
  \end{center}
\end{figure}

% GeSi CORE
\begin{figure}[t]
  \begin{center}
  \begin{tabular}{c c}
    \begin{tabular}{c c c }
    \hline
       NW Model & $(111)_\perp$ & $(100)_\perp$\\
    \hline
       \textbf{3\_3} &
       \begin{minipage}[h]{0.14\textwidth}
          \includegraphics[trim = 40mm 40mm 50mm 40mm, clip, width=1.\textwidth]{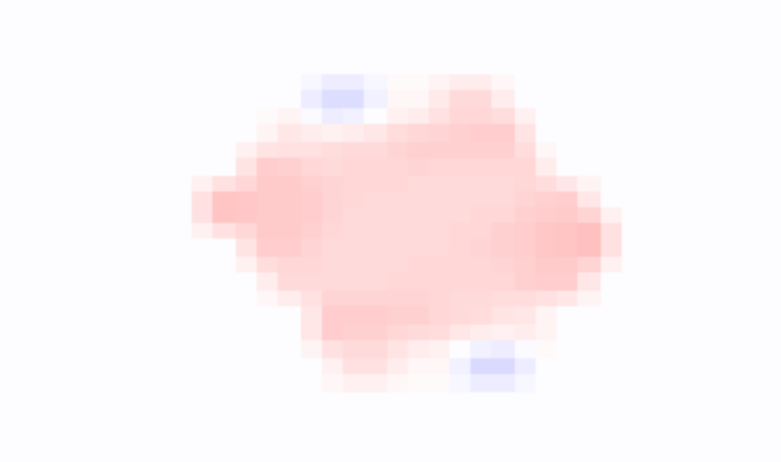}
        \end{minipage}&
       \begin{minipage}[h]{0.14\textwidth}
          \includegraphics[trim = 40mm 40mm 50mm 40mm, clip, width=1.\textwidth]{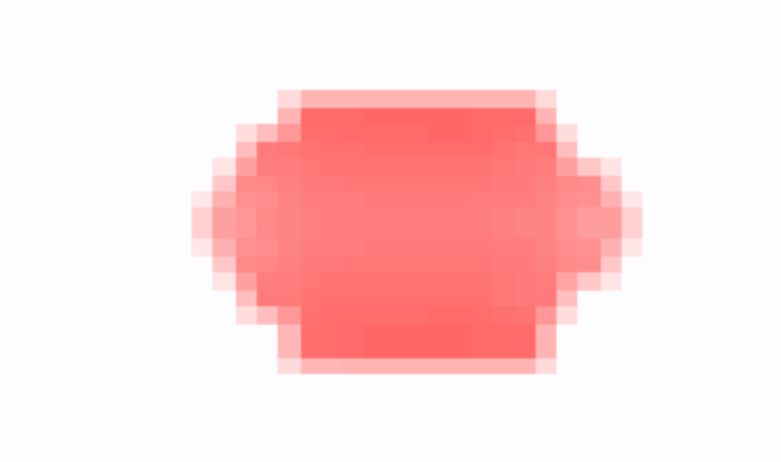}
        \end{minipage}\\
        \textbf{3\_5} &
        \begin{minipage}[h]{0.14\textwidth}
          \includegraphics[trim = 45mm 45mm 55mm 45mm, clip, width=1.\textwidth]{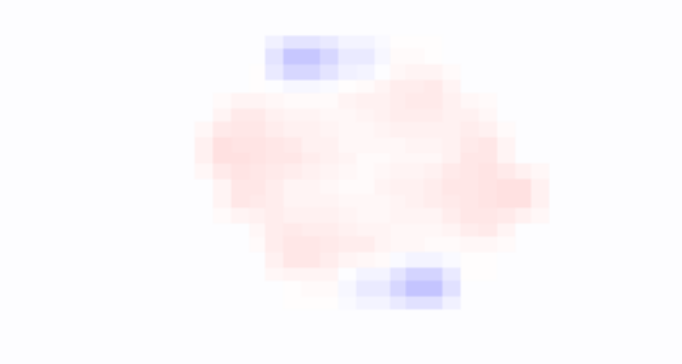}
        \end{minipage}&
       \begin{minipage}[h]{0.14\textwidth}
          \includegraphics[trim = 45mm 45mm 55mm 45mm, clip, width=1.\textwidth]{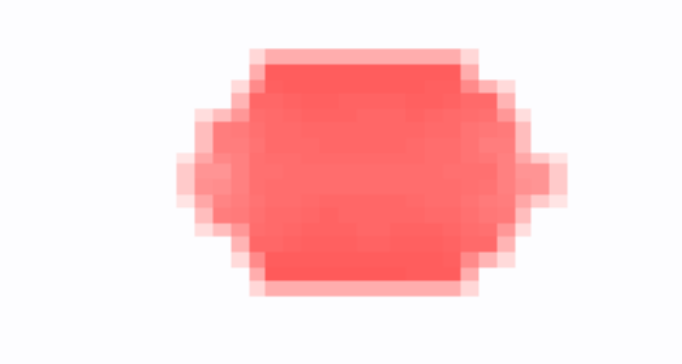}
        \end{minipage}\\
        \textbf{3\_7}&
         \begin{minipage}[h]{0.14\textwidth}
          \includegraphics[trim = 50mm 50mm 60mm 50mm, clip, width=1.\textwidth]{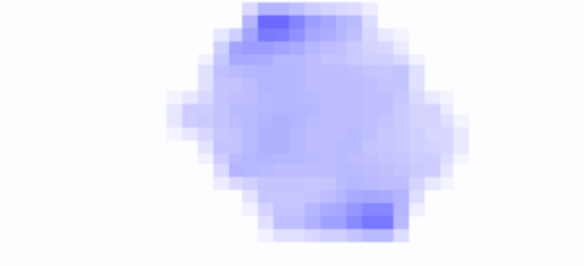}
        \end{minipage}&
       \begin{minipage}[h]{0.14\textwidth}
          \includegraphics[trim = 50mm 50mm 60mm 50mm, clip, width=1.\textwidth]{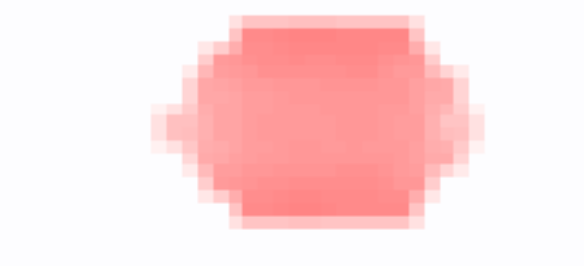}
        \end{minipage}\\
        \textbf{6\_3}&
         \begin{minipage}[h]{0.14\textwidth}
          \includegraphics[trim = 30mm 30mm 40mm 30mm, clip, width=1.\textwidth]{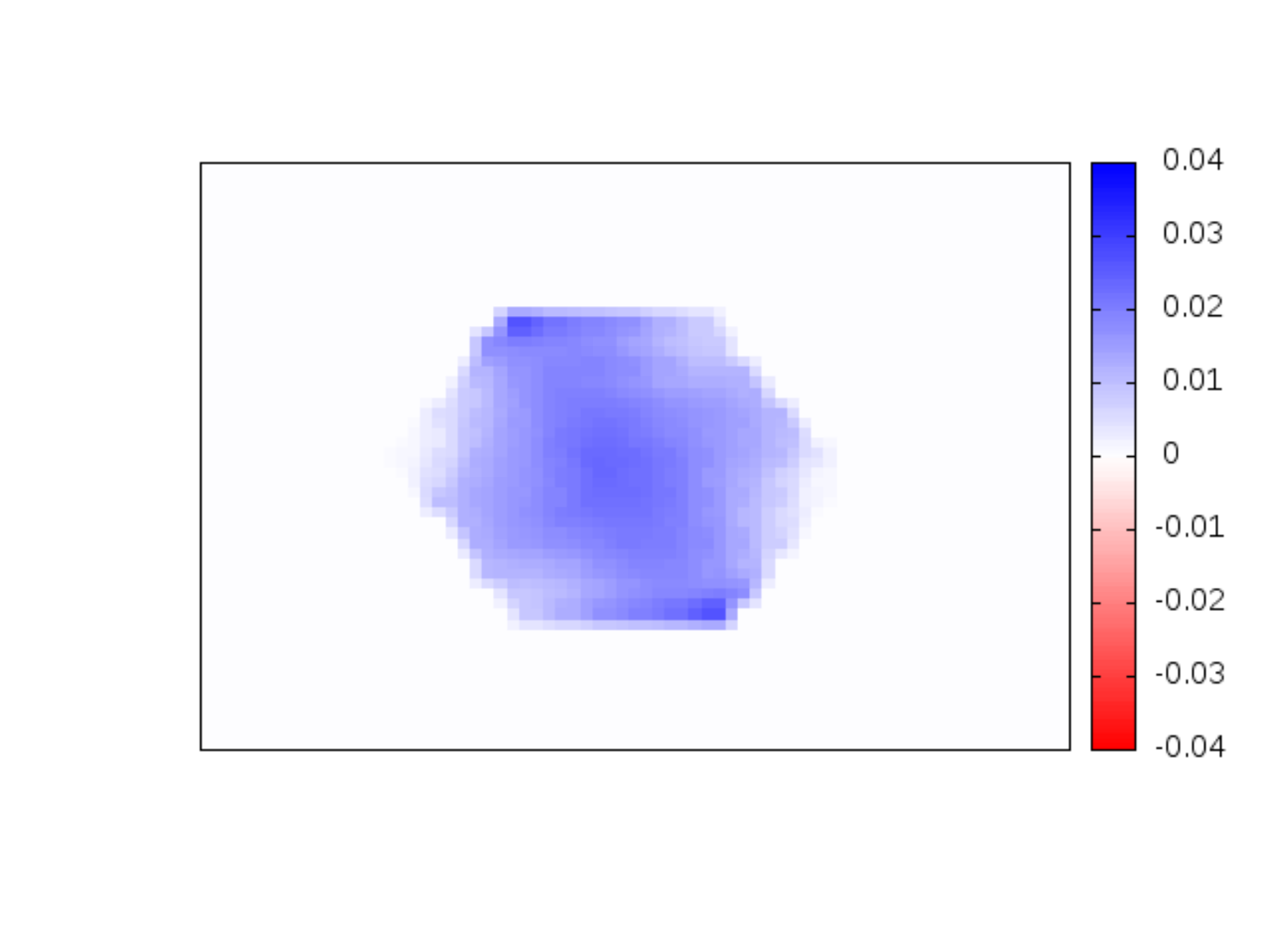}
        \end{minipage}&
       \begin{minipage}[h]{0.14\textwidth}
          \includegraphics[trim = 30mm 30mm 40mm 30mm, clip, width=1.\textwidth]{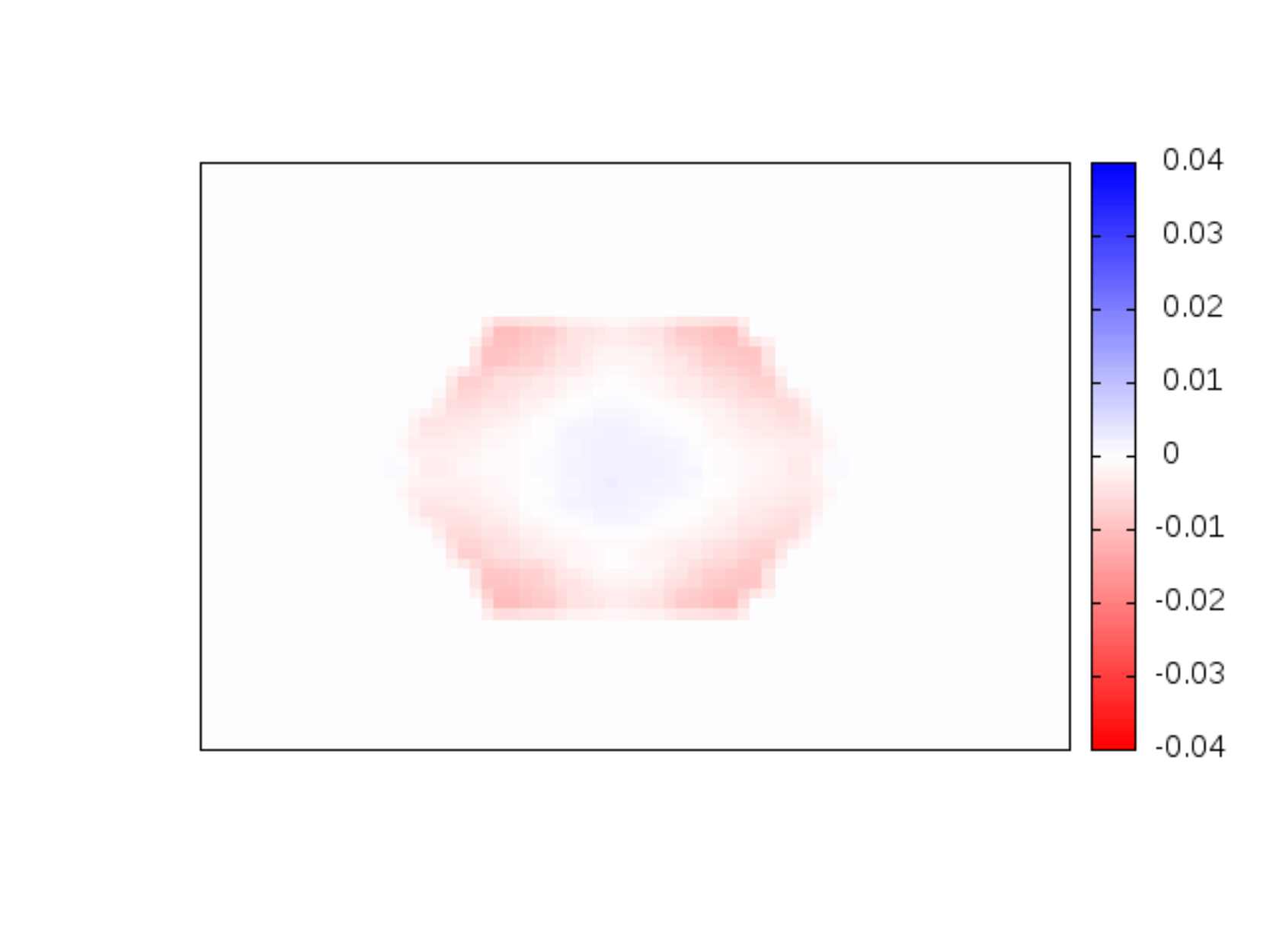}
        \end{minipage}\\
        \textbf{6\_5}&
         \begin{minipage}[h]{0.14\textwidth}
          \includegraphics[trim = 35mm 35mm 45mm 35mm, clip, width=1.\textwidth]{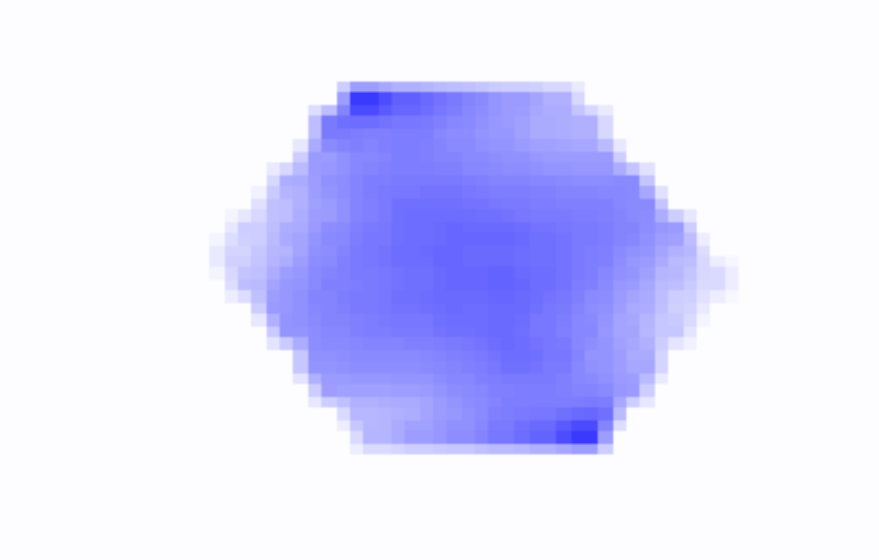}
        \end{minipage}&
       \begin{minipage}[h]{0.14\textwidth}
          \includegraphics[trim = 35mm 35mm 45mm 35mm, clip, width=1.\textwidth]{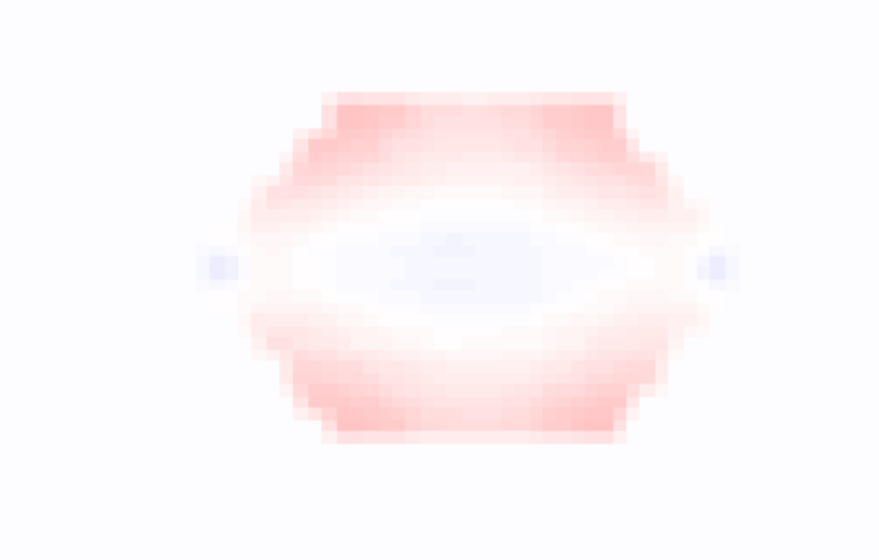}
        \end{minipage}\\
        \textbf{6\_7}&
        \begin{minipage}[h]{0.14\textwidth}
          \includegraphics[trim = 40mm 40mm 50mm 40mm, clip, width=1.\textwidth]{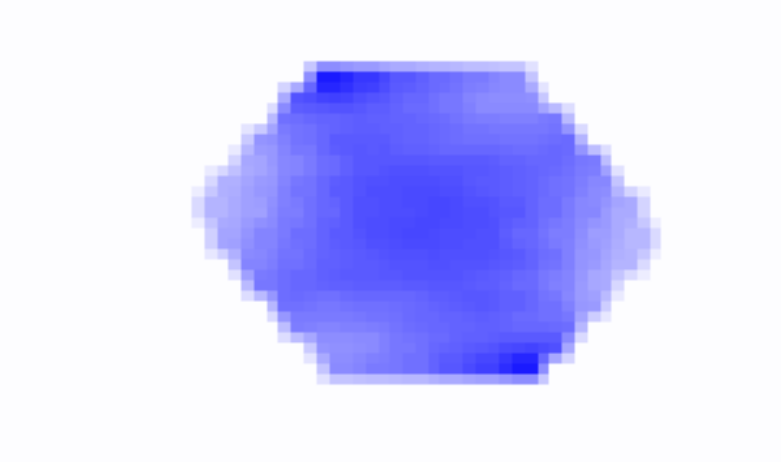}
        \end{minipage}&
       \begin{minipage}[h]{0.14\textwidth}
          \includegraphics[trim = 40mm 40mm 50mm 40mm, clip, width=1.\textwidth]{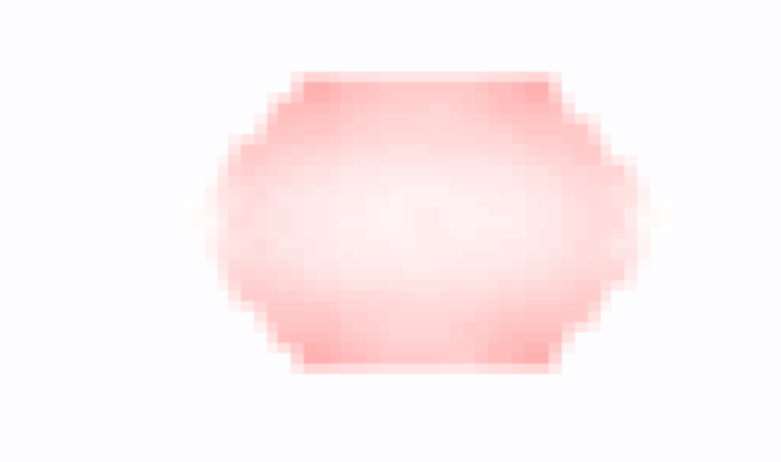}
        \end{minipage}\\
       \hline
    \end{tabular}
    &
    \begin{minipage}[h]{0.065\textwidth}
          \includegraphics[trim = 0mm 0mm 0mm 0mm, clip, width=1.\textwidth]{IMAGES/S3G7_shell_map_label.pdf}
        \end{minipage}\\
    \end{tabular}
    \caption{Average bond strain map for the cross-section of the Ge core of the GeSi-NWs. Maps for
the $(111)_\perp$ and $(100)_\perp$ labelled bonds are shown, with extension illustrated in blue and compression in red. }
     \label{core_bonds_GeSi}
  \end{center}
\end{figure}

Examining the strain maps for the Ge-core Si-shell nanowires, shown in Fig.~\ref{shell_bonds_GeSi} and Fig.~\ref{core_bonds_GeSi}, we can see that the basic behaviour has notable similarities to the SiGe case: the Si is under tensile stress, and is relatively isotropic, while the Ge is anisotropic, showing both tension and compression. 
There are significant differences, however, in both shell and core.  The Si shell shows smaller extension, particularly for the 3-layer core, which is often less than 1\% away from the (100) surface reconstruction.  The radial freedom has allowed more variation in the strain patterns to emerge. The $(111)_\perp$ bonds suffer the most significant extension when parallel to the closest (111) surface, i.e. in the second and fourth quadrants. The $(100)_\perp$ bonds have the largest extension at the interface between the two (111) Si surfaces, and the two (111) Ge surfaces, as well as at the reconstructed (100) surface with Si-Si dimers.

The Ge core shows less anisotropy than the Ge shell in the SiGe NWs, with a strong dependence on the thickness of the Si shell: increasing shell thickness leads to an increase in the tension (or equivalently decrease in compression) for the $(111)_\perp$ bonds.  This may seem counter-intuitive, as a thicker shell might be expected to lead to it having more influence, but reflects the increasing tension along this direction in the shell.  The shell clearly has more influence for the smaller core, seen particularly in the $(111)_\perp$ bonds.
The anisotropy of the Ge is still considerable, with the $(100)_\perp$ bonds generally compressed, while the $(111)_\perp$ bonds vary with NW size and location, showing particular variation at the interface between (111) and (100) surfaces.  As with the SiGe nanowires, we see that the germanium shows more anisotropy than the silicon.  It is clear that a careful, detailed \emph{ab initio} calculation is needed to describe the structure of nanowires properly: while simple extrapolation can approximate the axial lattice constant, the detailed structure depends sensitively on the details of the nanowire structure.

\subsection{Conclusion}

Using linear scaling DFT calculations, we have studied and compared the axial lattice parameters and intrinsic strain patterns of Si/Ge and Ge/Si core-shell nanowires with different core to shell ratios, with diameters in the range 4.9---10.2~nm.

We found that the axial lattice parameter calculated analytically using Vegard's empirical law gives a reasonable starting approximation to the axial lattice parameter, but that detailed DFT simulations are needed to find the correct values.  In some cases, the error in lattice parameter from Vegard's law was as large as 1\%. It is not surprising that a simple, empirical law based on solid solutions should only be approximately correct for these highly structured, anisotropic nanowires.  Our DFT calculations show that increasing the Si content leads to a reduction in the axial lattice parameter, towards the value of bulk Si.  With one exception, increasing the Ge content results in an increase in the axial lattice parameter towards the value of bulk Ge, though in all cases the behaviour does not follow the proportions of the constituents in a simple, linear fashion.

The lattice constant of bulk Ge is greater than that of bulk Si by 4\% and this generates an intrinsic strain in Ge and Si bonds in the nanowires. In all Si/Ge and Ge/Si nanowires we have studied, the Si component shows generally isotropic expansion, while the Ge component shows complex, bond-direction and nanowire-size-dependent tensile and compressive strain patterns. Given that the core does not have the same radial freedom as the shell, it is interesting that the Si component behaves consistently, regardless of whether it is placed as the core or the shell. The differences in the elastic properties of silicon and germanium may play a role in this behaviour, reflecting the different bond strengths in the two materials.  The varying strain within core and shell will have a variable effect on the electronic structure\cite{Leu:2008pj,Niquet:2012ux}, requiring careful \emph{ab initio} modelling to quantify the details.

The most highly strained and anisotropically strained areas were observed near the surfaces and heterojunction interface, and structural deformation may be more likely in these areas. Areas of sharp transition from tensile to compressive strain can be seen near the core-shell inteface of certain nanowires (e.g. Si/Ge 6\_5 and 6\_7 structures, Ge/Si 3\_3 and 3\_5). In such cases, the diffusion of Ge into the strained Si layer or Si into strained Ge layers is likely to be enhanced over standard rates, and as this will change the composition of the nanowires, could pose performance issues in core-shell nanowires.  The strain due to both surface reconstruction and relaxation at the surfaces will affect the local chemical reactivity.

In summary, we have shown that linear scaling DFT is both possible and necessary for systems of this size, and have produced a detailed study of the strain patterns in silicon-germanium core-shell nanowires.

\begin{acknowledgments}
We thank Dr. N. Fukata for useful conversations.  TM acknowledges support from JSPS Grant-in-Aid for Scientific Research: Grant Number 26246021, Japan.
\end{acknowledgments}

\bibliography{NW}

\end{document}